\documentclass[fleqn,usenatbib]{mnras}

\usepackage{newtxtext,newtxmath}
\usepackage[T1]{fontenc}
\usepackage{ae,aecompl}

\usepackage{graphicx}	
\usepackage{amsmath}	
\usepackage{amssymb}	
\usepackage{pdflscape}	
\usepackage{longtable}
\usepackage{threeparttable} 
\usepackage{upgreek} 
\usepackage{booktabs}
\usepackage{caption}
\usepackage{multicol} 

\newcommand{\HI}{H~{\sc i}}
\newcommand\Ms{$M_\ast$}
\newcommand\Msolar{$\mathrm{M_\odot}$}
\newcommand\Mh{$M_{{H {\sc I}}}$}

\title[LIERS]{\HI\ galaxies with little star formation: an abundance of LIERs.}

\author[V. Parkash et al.]{
Vaishali Parkash,$^{1}$\thanks{E-mail: vaishali.parkash@monash.edu}
Michael J.I. Brown,$^{1}$
T. H. Jarrett$^{2}$
A. Fraser-McKelvie$^{3}$
M.E. Cluver$^{4,5}$
\\
$^{1}$ School of Physics and Astronomy, Monash Centre for Astrophysics (MoCA), Monash University, Clayton, Victoria 3800, Australia \\
$^{2}$ Astrophysics, Cosmology and Gravity Centre (ACGC), Astronomy Department, University of Cape Town, Private Bag X3, Rondebosch 7701, South Africa \\
$^{3}$ School of Physics and Astronomy, University of Nottingham, University Park, Nottingham, NG7 2RD, UK\\
$^{4}$ Centre for Astrophysics and Supercomputing, Swinburne University of Technology, John Street, Hawthorn, 3122, Australia\\
$^{5}$ Department of Physics and Astronomy, University of the Western Cape, Robert Sobukwe Road, Bellville, 7535, South Africa
}

\date{Accepted 2019 February 23. Received  2019 February 18; in original form 2018 December 11.}

\pubyear{2018}

\begin{document}
\label{firstpage}
\pagerange{\pageref{firstpage}--\pageref{lastpage}}
\maketitle

\begin{abstract}
We present a sample of 91 \HI\ galaxies with little or no star formation, and discuss the analysis of the integral field unit (IFU) spectra of 28 of these galaxies. We identified \HI\ galaxies from the \HI\ Parkes All-Sky Survey Catalog (HICAT) with Wide-field Infrared Survey Explorer (WISE) colours consistent with low specific star formation ($\mathrm{<~10^{-10.4}~yr^{-1}}$), and obtained optical IFU spectra with the Wide-Field Spectrograph (WiFeS). Visual inspection of the PanSTARRS, Dark Energy Survey, and Carnegie-Irvine imaging of 62 galaxies reveals that at least 32 galaxies in the sample have low levels of star formation, primarily in arms/rings. New IFU spectra of 28 of these galaxies reveals 3 galaxies with central star formation, 1 galaxy with low-ionisation nuclear emission-line regions (LINERs), 20 with extended low-ionisation emission-line regions (LIERs) and 4 with high excitation Seyfert (Sy) emission. From the spectroscopic analysis of \HI\ selected galaxies with little star formation, we conclude that 75\% of this population are LINERs/LIERs.
\end{abstract}




\section{Introduction}

Neutral atomic hydrogen (\HI) gas plays a vital role in the evolution and growth of galaxies, as \HI\ gas is the principal cool gas reservoir within galaxies from which molecular hydrogen, $\mathrm{H_2}$, can form and then produce stars. Local galaxies fall into two main populations: optically-red, dead spheroids, and blue, young, and star-forming disc galaxies \citep{tully82, blanton03, baldry04, schiminovich07}. This bimodal distribution suggests that blue disc galaxies grow via star formation, quench their star formation and often transition to early-type morphology when their \HI\ gas reservoirs are heated or removed  \citep[e.g, ][]{larson80, bell04, martin12} by a diverse set of mechanisms that operate via quenching or regulation of the gas supply \citep[e.g,][]{keres05,hopkins08,martin07}. Understanding the mechanisms through which galaxies deplete their \HI\ gas will contribute to our understanding of the evolution of galaxies and their different properties. 

Large ``blind" surveys of \HI\ have allowed us to study and measure the correlations between star formation, stellar mass and \HI\ content of galaxies in the local Universe. Unsurprisingly, the \HI\ content of galaxies does correlate with stellar mass and star formation rate \citep[SFR; e.g.,][]{haynes84, doyle06, catinella10, cluver10, huang12, denes14, parkash18}. \citet{doyle06} found that the \HI\ masses of galaxies increase as a function SFR, with power-law indices of $0.6$ and $0.4$ for infrared (IR)-derived SFRs and 1.4 GHz-derived SFRs, respectively. Using the \HI\ sample from \citet{parkash18} and a contemporary WISE W3 SFR calibration, we estimate \HI\ mass increases with SFR as a power-law with an index of 0.45. However, we also see individual galaxies to have a large scatter about this relation, with a $1\sigma$ scatter of 0.35~dex.

The observed relationship between \HI\ gas mass and SFR is predicted by hydrodynamical simulations \citep[e.g.,][]{vogelsberger14, lagos16}. In order to maintain star formation, galaxies replenish their gas content from multiple sources \citep[e.g.,][]{benson10,somerville15} including accreting gas from the intergalactic medium \citep[IGM;][]{binney04, keres05, dekel09}, from their halos via a galactic fountain mechanism \citep{fraternali02, fraternali11}, or from gas-rich mergers \citep{sancisi08}. The gas then settles into a disk in the galaxy and will collapse to create molecular gas and then stars when the disk becomes sufficiently dense. This interplay between gas inflow and star formation is essential for reproducing observed scaling relations between properties of galaxies and gives rise to well-known correlations such as the Kennicutt-Schmidt law \citep{kennicutt98}. However, the relationship between SFR-\HI\ breaks down when the \HI\ column density \citep{bigiel10} is so low that the angular momentum prevents the gas from collapsing \citep{kim13,obreschkow16}. More recent work of \citet{bacchini18} \\ shows that the Kennicutt-Schmidt law does not break after the conversion to  \textit{volume densities}, suggesting that the break in surface-based laws is caused by flaring of the gas disc rather than an inefficiency of star formation at low surface densities.

Multiple groups have targeted galaxies that are outliers from scaling relations between SFR, stellar mass and \HI\ mass of galaxies, including \HI-excess galaxies \citep{lutz17, gereb16, gereb18} and elliptical galaxies with \HI\ gas \citep{oosterloo02, oosterloo10, serra12}, with the goal of understanding the mechanisms that regulate star formation. An archetype \HI-excess galaxy, GASS 3505, with \HI\ mass (\Mh)$=$ $10^{9.9}~{\rm M_\odot}$ and a surprisingly inefficient SFR of $\sim 0.1~{\rm M_\odot~yr^{-1}}$, was found to be associated with a 50~kpc gas ring. Evidence of recent star formation was observed in regions of high \HI\ column densities ($\mathrm{>~10^{20}~cm^{-2}}$), but not in regions of low \HI\ column densities, suggesting that low \HI\ column density is likely the main factor for the low SFR \citep{gereb16}. It has also been proposed that the high specific angular momentum of \HI-excess galaxies prevents the gas in these galaxies from collapsing inwards \citep{hallenbeck14, boissier16, obreschkow16, lutz17}. Some early-type galaxies (ETGs; ellipticals and lenticulars) have also been found to have \HI\ gas masses similar to spirals \citep[e.g,][]{knapp85, sadler02, oosterloo10, serra12}. However, while the \HI\ column density distribution of spirals is very broad, reaching up to $\mathrm{10^{21}~cm^{-2}}$, most \HI-rich ETGs have \HI\ column densities $\mathrm{<~5\times10^{20}~cm^{-2}}$ \citep{serra12, yildiz15}. ETGs have column densities that are similar to those observed in the outer regions of spirals where very little star formation occurs \citep{bigiel10}.

The aim of this work is to identify a sample of \HI\ galaxies with low SFRs and gain insight into why these galaxies are inefficient at converting their \HI\ into stars. We have compiled a sample of \HI\ galaxies with low or no star formation based on their WISE mid-IR photometry \citep{parkash18} and obtained integral field unit (IFU) observations for a subsample. The paper is arranged as follows: Section~\ref{sec:sample} defines the sample; Section~\ref{sec:obs} details the IFU observation and data; Section~\ref{sec:results}~and~\ref{sec:lier} describes the galaxies with IFS; Section~\ref{sec:dis} discusses the results; and Section~\ref{sec:conclusion} summarises our work. All magnitudes are in the Vega system. The cosmology applied in this paper is $\mathrm{H_0 ~=~70~km~s^{-1},~\Omega_M~=~0.3,~and~\Omega_\Lambda~=~0.7}$.


\section{Sample}\label{sec:sample}

\subsection{Sample Selection}

We selected this sample from \HI\ Parkes All-Sky Survey (HIPASS) Catalogue \citep[HICAT;][]{meyer04} and new Wide-field Infrared Survey Explorer (WISE) photometry \citep{wright10, parkash18}. HICAT comprises of 4315 \HI\ sources below a declination $\delta$ of +2 $\degr$ detected by Parkes 64 m radio telescope using a 21 cm multibeam receiver. The catalogue is 99\% complete at a peak flux of $84~{\rm mJy}$ and an integrated flux of $9.4~{\rm Jy~km~s^{-1}}$ \citep{zwaan04}.

WISE, launched on December 2009, mapped the entire sky in the W1, W2, W3, and W4 infrared bands, corresponding to 3.4, 4.6, 12, and $22~{\rm \mu m}$ respectively \citep{wright10}. The four WISE bands achieved point source sensitivities in Vega magnitudes of 16.5, 15.5, 11.2, and 7.9, respectively, with W4 being approximately two orders of magnitude more sensitive than the Infrared Astronomical Satellite (IRAS) equivalent band. The W1 and W2 bands are dominated by light from K- and M-type giant stars and are good tracers of the underlying stellar mass of a galaxy \citep[e.g.,][]{meidt12, cluver14}. The W3 band is a star formation rate indicator that is dominated by emission from polycyclic aromatic hydrocarbon (PAH) features, warm dust, and silicate absorption \citep[e.g.,][]{calzetti07,cluver17}. The WISE mid-infrared bands are a powerful tool to measure the stellar mass and star formation rates of galaxies at low redshifts with minimal extinction. 

In \citet{parkash18}, we constructed new WISE image mosaics and measured improved photometry for 2831 HICAT galaxies to mitigate limitations in the default ALLWISE photometry. ALLWISE photometry \citep{cutri13} is optimised for point sources \citep[e.g.][]{jarrett13} and therefore is not ideal for resolved sources such as HICAT galaxies. Also, the elliptical apertures (gmag) in the WISE catalogs for 2MASS Extended Sources \citep[2MXSC;][]{jarrett00} miss a significant fraction of the flux. For $W1\sim 10$ mag galaxies, the new W1 and W3 magnitudes are on average systematically brighter by 0.6 and 0.8 mags (respectively) than the default pipeline magnitudes, while larger offsets can occur for fainter galaxies \citep{parkash18}. For further details about the new photometry, we refer the reader to \citet{jarrett13} and \citet{parkash18}. 

To select \HI\ sources with little or no star formation we implemented the following selection criteria:
\begin{enumerate}
 \item A HICAT source with observed W2-W3 colour $<$ 2.0, which corresponds to galaxies with a specific star formation rate (sSFR) $<$~10$^{-10.4}$~$\mathrm{yr^{-1}}$, based on the W3 SFR calibration of \citet{brown17} and W1 stellar mass calibration of \citep{cluver14}.
 \item A stellar mass cut $\ge$ $\rm{10^{10}}$ \Msolar\ to exclude low-mass dwarf galaxies. Stellar masses are estimated from the stellar mass-to-light ($M_\ast/L_{W1_\odot}$) ratio relation of \citet{cluver14} where the W1-W2 colour dependence takes into account the morphological dependence on the M/L (see Appendix \ref{sec:sample_app} for further details). Dwarf galaxies are optically thin systems in terms of their dust properties, therefore the UV light produced by the young and massive stellar populations escapes the galaxy, causing their W2-W3 colour to be similar to passive galaxies \citep[e.g.,][]{jarrett17}.
 \item Galaxies at least 10 degrees away from the Galactic plane, as high Galactic dust extinction and a high density of foreground stars complicates the photometric measurements.
 \item \HI\ sources that are not comprised of multiple galaxies. 
 \end{enumerate}

\begin{figure}
\centering
\includegraphics[width=\columnwidth]{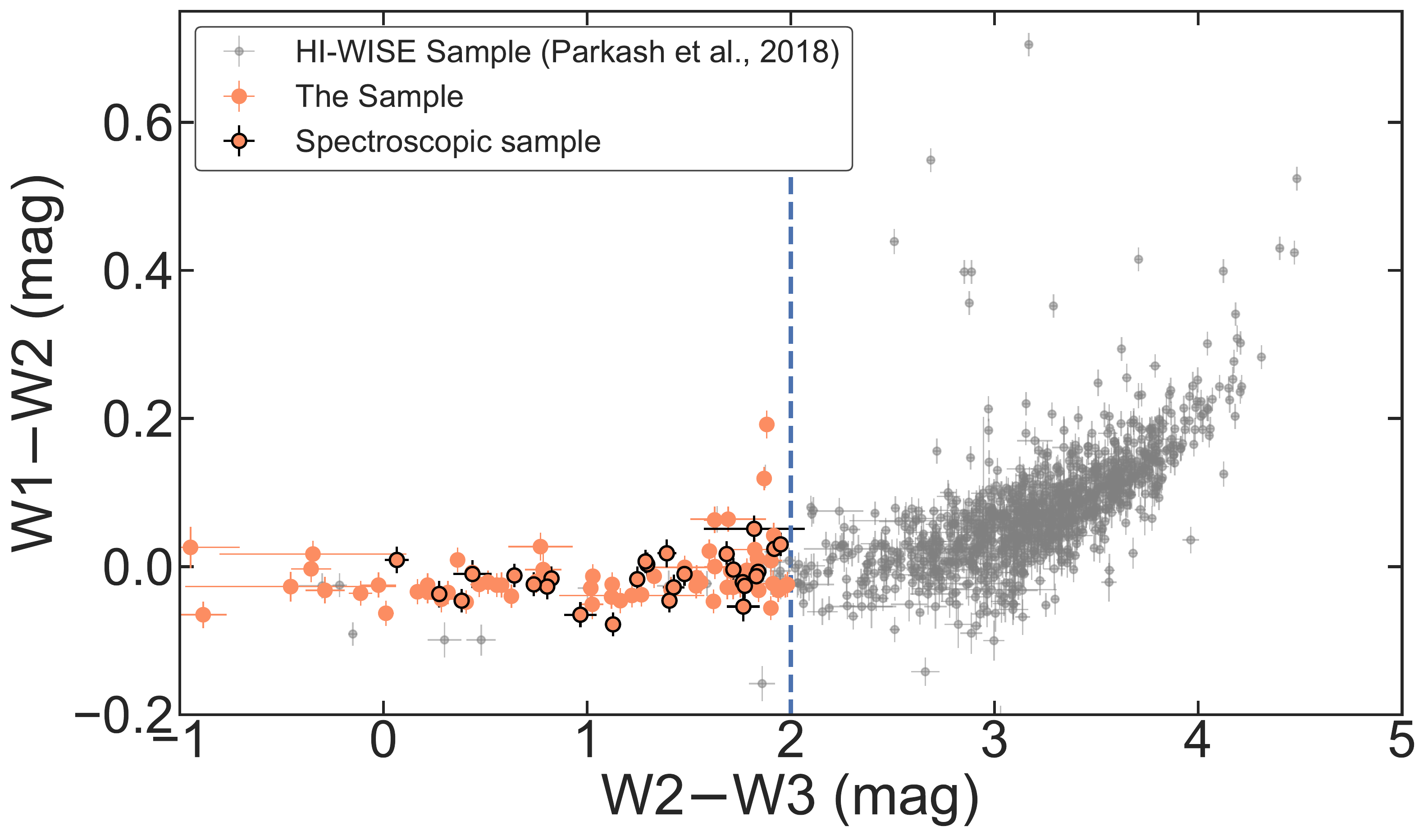}
\caption{WISE mid-infrared colours of the \HI-WISE sample and the \HI\ sample with little to no star formation. The vertical blue line denotes the division between spheroids with little star formation (W2-W3 $<$ 2) and star-forming, intermediate, or late-type disk galaxies (W2-W3 $>$ 2) \citep[e.g. ][]{jarrett17}. Galaxies from this \HI\ sample with follow-up spectroscopic data are circled in black. Though \HI-WISE is dominated by star-forming, intermediate, or late-type disk galaxies, there exist \HI\ galaxies with relative passive WISE W2-W3 colours. We have selected these outliers to study the proprieties of \HI\ galaxies with little to no star activity. }
\label{fig:colour}
\end{figure}

In addition to these cuts, the optical images of each galaxy are inspected and galaxies in crowded star fields or near bright foreground stars are removed from the sample. The final sample contains 91 $\mathrm{z < 0.04}$ galaxies and we illustrate their WISE colours in Figure \ref{fig:colour}, along with galaxies from \citet{parkash18} for comparison. 

\subsection{The Sample in Context}

\begin{figure}
\centering
\includegraphics[width=0.89\columnwidth]{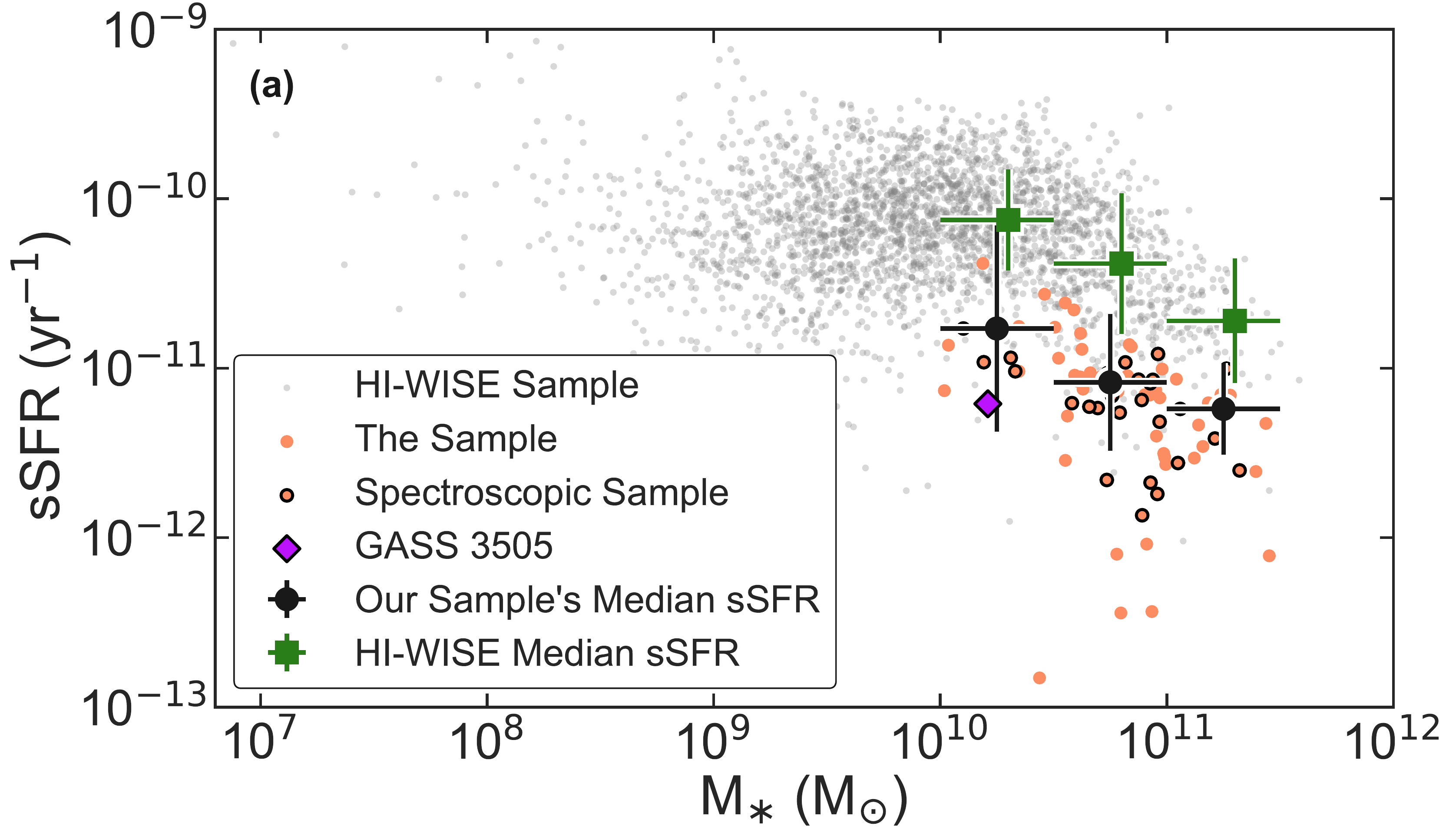}
\includegraphics[width=0.89\columnwidth]{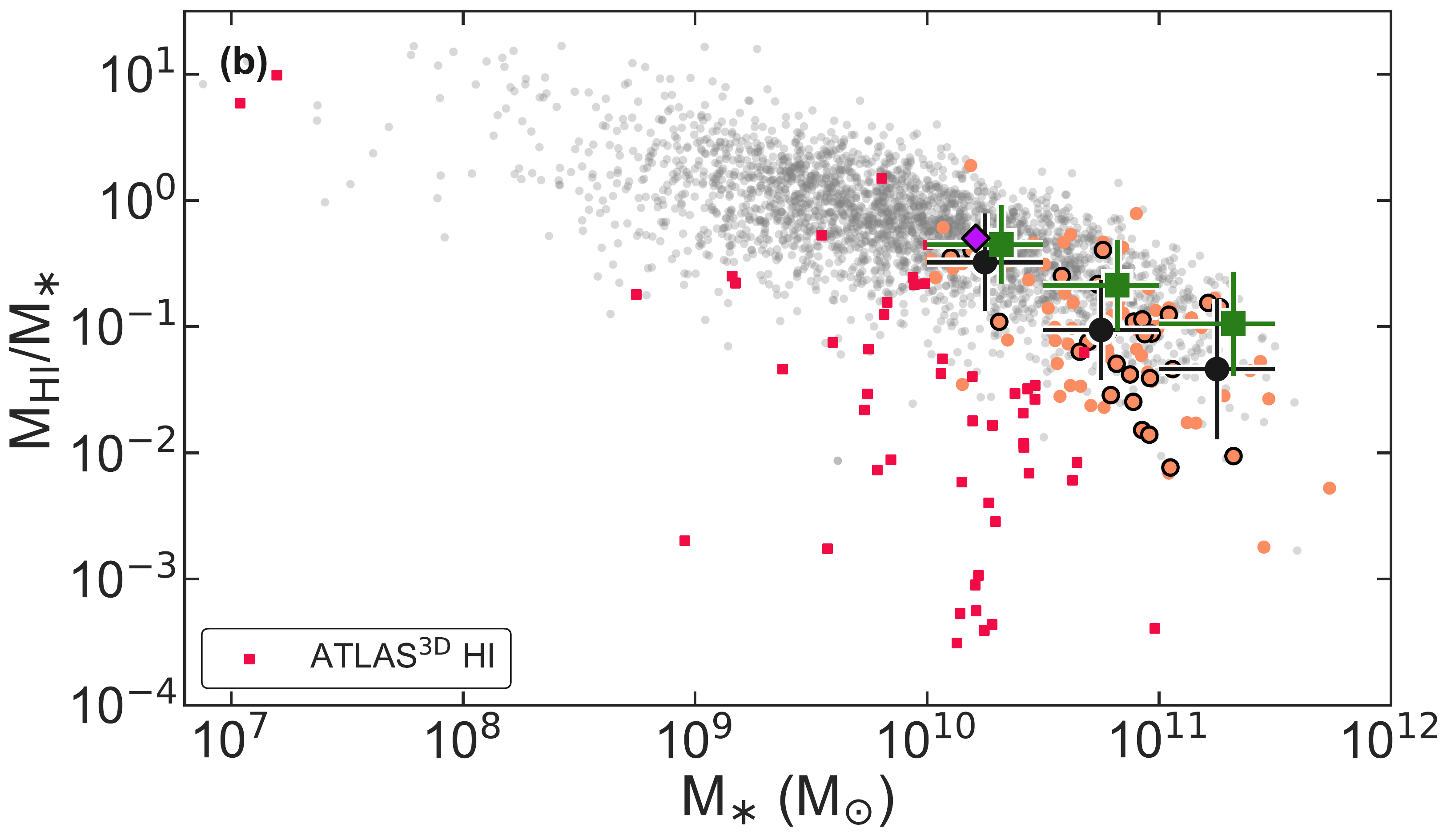}
\caption{The stellar and gas properties of the \HI\ galaxies in the sample compared to the \HI-WISE sample: a. sSFR (SFR/\Ms) versus stellar mass and b. \HI\ mass fraction (\Mh/\Ms) versus stellar mass. Median values are shown for the sample (black circles) and the \HI-WISE sample (green squares). Medians for the \HI-WISE sample were estimated using only galaxies that met the last three sample selection criteria. For comparison, we also plot the \HI\ and stellar properties of well studied galaxy GASS 3503 \citep[purple diamond; ][]{gereb16} and the ETGs from $\mathrm{ATLAS^{3D}}$ \HI\ survey \citep[pink; ][]{serra12}. By construction, the sample has a lower sSFR than the \HI-WISE sample. On average, the \HI\ gas fraction is lower than those of typical \HI\ galaxies, but a significant fraction of galaxies have \HI\ masses similar to those of typical star-forming galaxies. GASS 3503, a well-studied example of this class of galaxy, falls within the range of the sample, albeit with relatively low sSFR and high \HI\ mass. GASS 3503 and some ETGs have \HI\ content comparable to the galaxy sample emphasising that these galaxies are among the least star-forming \HI\ galaxies. }
\label{fig:SFR_MASS}
\end{figure}

In Figure \ref{fig:SFR_MASS} we compare the sSFR and \HI\ gas fraction (defined as \Mh/\Ms) distribution of the sample to the larger \HI-WISE sample of \citet{parkash18}.  The estimated median sSFR and \HI\ gas fraction for stellar mass bins with a width of log(\Ms) $=$ 0.5 and 1$\mathrm{\sigma}$ (determined using the range encompassing 68\% of the data) for both samples are listed in Table \ref{table:med}. The medians for the \HI-WISE sample are only determined using galaxies that meet the last three selection criteria (ii-iv). The sample falls below the \HI-WISE sample and the star-forming main sequence \citep[MS;][]{noeske07a, rodighiero11,wuyts11} by construction. Also, while the \HI\ gas fraction for the sample is systematically lower than those of most comparable spiral galaxies, there is a broad spread and significant overlap. Therefore the low star formation rate of these galaxies cannot be explained by depletion of \HI\ gas alone.
 
In Figure \ref{fig:SFR_MASS} we plot GASS 3505 (purple diamond), which is a well-studied red sequence galaxy with a relatively high \HI\ mass \citep{gereb16}. The \HI\ mass and low SFR of GASS 3505 is comparable to some of our higher \HI\ mass galaxies. This further emphasises that the sample galaxies are among the least star-forming \HI-rich galaxies.

We also compare this sample to the $\mathrm{ATLAS^{3D}}$ \HI\ survey \citep{serra12}, which undertook a 21-cm survey of 166 early-type galaxies with the Westerbork Synthesis Radio Telescope (WSRT). Only 52 (32\%) of the \citet{serra12} early-type galaxies had \HI\ detections, with \HI\ masses ranging from 10$^{6.26}$ to 10$^{9.98}$ \Msolar. In Figure~\ref{fig:SFR_MASS} we compare the \HI\ content and stellar properties of $\mathrm{ATLAS^{3D}}$ and the sample, using NASA-Sloan Atlas (NSA) masses for the $\mathrm{ATLAS^{3D}}$ galaxies \citep{blanton07}. The \HI\ gas fraction is on average lower for the $\mathrm{ATLAS^{3D}}$ ETG \HI\ sample than our HIPASS selected sample, as $\mathrm{ATLAS^{3D}}$ \HI\ observations were a targeted survey of individual galaxies and could thus push deeper than the blind HIPASS survey. That said, there are ATLAS$^{\mathrm{3D}}$ ETGs with \HI\ masses comparable to those of galaxies from the sample and typical star-forming spiral galaxies.

\begin{table}
\center
\caption{Median sSFR, \HI\ gas fraction and the $1\sigma$ (68\%) scatter for each stellar mass bin.}
\label{table:med}
\resizebox{\columnwidth}{!}{
\begin{tabular}{ccccc}
\toprule
& \multicolumn{2}{c|}{This Work}&\multicolumn{2}{c|}{\HI-WISE Sample}\\
\midrule
log(\Ms) &  log sSFR & log(\Mh/\Ms) & log sSFR & log(\Mh/\Ms) \\
\Msolar &  yr$^{-1}$ & &  yr$^{-1}$ &  \\\hline
10.25 & -10.8$\pm$0.6 & -0.5$\pm$0.4 & -10.1$\pm$0.3 & -0.3$\pm$0.3\\ 
10.75 & -11.1$\pm$0.4 & -1.0$\pm$0.4 & -10.4$\pm$0.4 & -0.7$\pm$0.4\\
11.25 & -11.2$\pm$0.3 & -1.3$\pm$0.6 & -10.7$\pm$0.4 & -1.0$\pm$0.4\\
\bottomrule
\end{tabular}
}
\end{table}

\begin{figure*}
\includegraphics[width=0.9\textwidth]{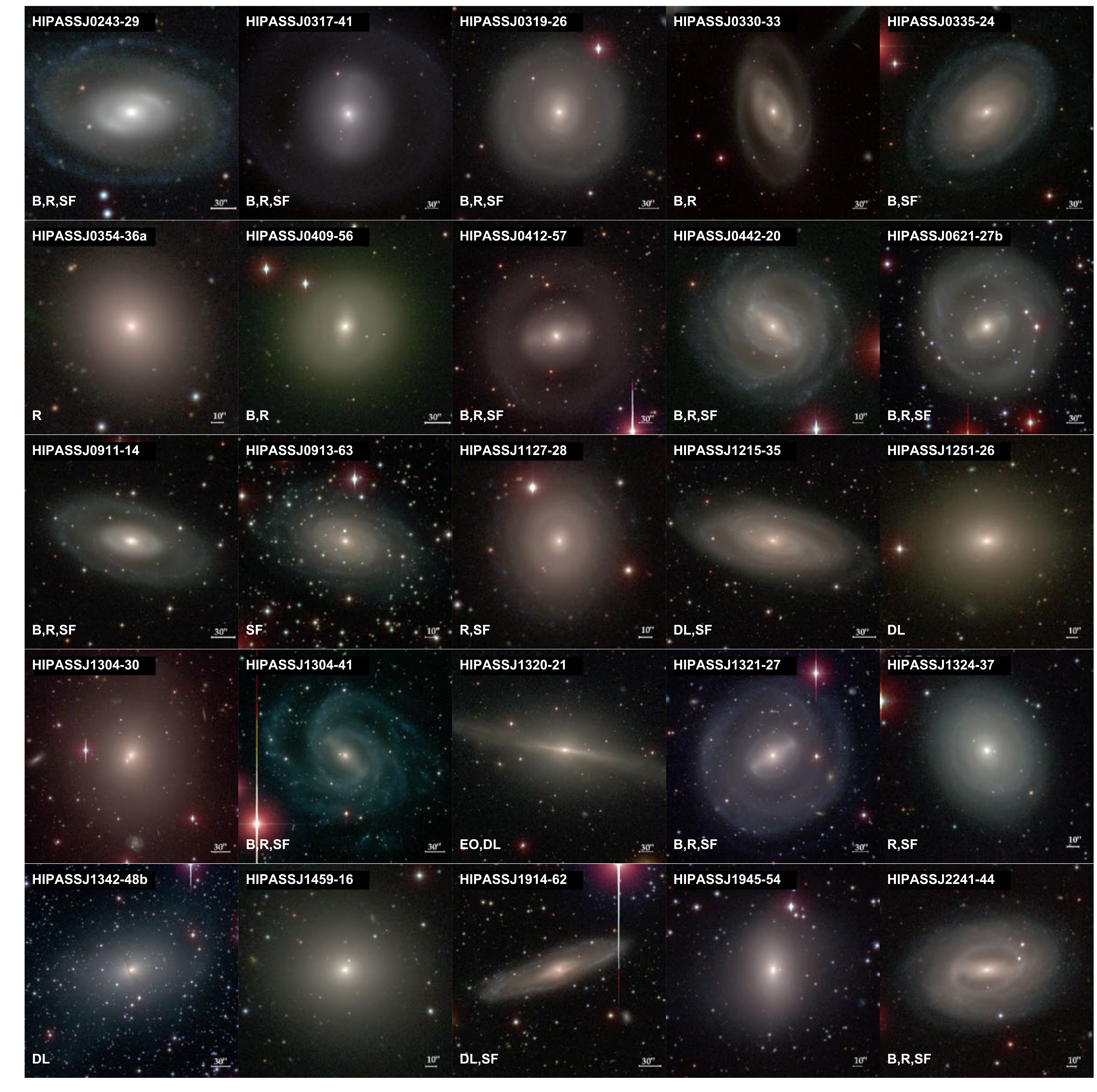}
\caption{CGS cutout gri images \citep[CGS;][]{ho11} of 25 galaxies in the sample. The bottom left of each postage stamp includes notes on the optical features of the galaxy such as bars (B), rings (R), star formation in the outer region (SF), edge-on (EO) and dust lanes (DL). Out of these 25 galaxies, 15 exhibit either bar-like, ring-like features or both and 16 galaxies are passive at the centre but show signs of recent star formation in the outskirts. On average, these star-forming rings are $\mathrm{\sim1^{\prime}}$ from the centre of the galaxy, which at the sample's average redshift of$\mathrm{\sim0.01}$ corresponds to a radius of $\mathrm{\sim10}$ kpc. For this work, we have only observed the central $\mathrm{25^{\prime\prime} \times 38^{\prime\prime} }$ region of 28 galaxies with Wide Field Spectrograph \citep[WiFeS,][]{dopita07, dopita10} integral field unit (IFU) and therefore do not have observe the faint outer star-forming regions.}
\label{fig:CGS}
\end{figure*}

To examine the morphologies of the sample and search for low level star formation, we use images from Carnegie-Irvine Galaxy Survey \citep[CGS;][]{ho11}, Dark Energy Survey \citep[DES; ][]{flaugher15,abbott18,morganson18}, and Panoramic Survey Telescope and Rapid Response System \citep[Pan-STARRS; ][]{flewelling16}. In Figure \ref{fig:CGS} we present example postage stamps images from CGS. Of the 91 galaxies in the sample, 62 galaxies have deep optical images from at least one of these three surveys. By visually inspecting the deep optical images of the 62 galaxies, we find that 35 galaxies exhibit bars and/or rings. We also find that 32 of the 62 galaxies show clear evidence of recent star formation, including blue spiral arms or rings from the deep optical images at large radii (on average $\mathrm{\sim1^{\prime}}$, or $\sim$10 kpc at $z=0.01$, but up to $\sim$30 kpc). Excluding those galaxies that show clear or tentative evidence of star formation in deep optical or GALEX ultraviolet images, we find just 9 galaxies (including HIPASSJ1304-30 and HIPASSJ1459-16) that could potentially be passive. Table \ref{table:full_sample} provides notes on the morphologies and evidence for star formation for individual galaxies, along with redshifts, stellar masses and \HI\ masses.

\section{Spectroscopic Observations and Data Reduction}\label{sec:obs}
\subsection{WiFeS Observations}
We observed 28 galaxies from the sample using the Wide Field Spectrograph \citep[WiFeS,][]{dopita07, dopita10} integral field unit (IFU) on the Australian National University's 2.3 m telescope at Siding Spring Observatory. We will refer to these 28 galaxies as the spectroscopic sample. WiFeS IFU has a 25$\mathrm{^{\prime\prime}}$ $\times$ 38$\mathrm{^{\prime\prime}}$ field of view and we observed each galaxy using the B3000 and R3000 gratings along with the RT560 dichroic, providing wavelength coverage from 3500-9200 \AA\ and spectral resolution of $\sigma$ $\sim$ 100 km s$^{-1}$. The observations were taken in nod-and-shuffle mode to reliably subtract the skylines. The WiFeS data were reduced using a Python-based data reduction pipeline, PyWiFeS, developed by \citet{childress14}. Table \ref{tab:obs} lists the galaxies observed with WiFeS. Targets were prioritised based mainly upon observability, though galaxies that did not show any evidence of observable nuclear star formation in their optical and or UV images were preferred. 

\begin{landscape}
\begin{table}
\begin{threeparttable}
\caption{WiFeS observations of 28 galaxies in the sample.} \label{tab:obs}
\begin{tabular}{lllllllllll}
 HIPASS ID & Name & RA & DEC & z & Obs Date & $\mathrm{D_{4000}^a}$ & Age$^b$& log$_{10}$([N {\sc ii}] /H$_\alpha$)$^c$ & log$_{10}$([O {\sc iii}]/H$_\beta$)$^c$  & Spectral$^d$   \\
&  & (J2000) & (J2000) &  &  &  & (Gyr) & & & Type   \\ \hline
0150-47  & ESO 245-G006           &01:50:28.7 & -47:09:57 & 0.0207 & 9 July 2018     & 1.85 &	4.32 & 0.082    &  0.1890  &  LIER	   \\
0154-00b & UGC 01382              &01:54:41.0 & -00:08:36 & 0.0192 & 22 July 2017    & 1.94 &	6.43 & -0.075   &  0.266   &  LIER	   \\
0243-29  & NGC 1079               &02:43:44.3 & -29:00:12 & 0.0048 & 8 August 2018   & 1.60 &	1.87 & -0.314   &  -0.125  &  LIER	   \\
0319-26  & NGC 1302               &03:19:51.2 & -26:03:38 & 0.0057 & 9 August 2018   & 1.79 &	3.12 & 0.034    &  0.462   &  LIER	   \\
0330-33  & NGC 1350               &03:31:08.1 & -33:37:43 & 0.0063 & 11 July 2018    & 1.88 &	4.99 & 0.141    &  0.541   &  LIER	   \\
0409-56  & NGC 1533               &04:09:51.8 & -56:07:06 & 0.0026 & 7 August 2018   & 1.96 &	6.97 & 0.302    &  0.403   &  LIER	   \\
0412-57  & NGC 1543               &04:12:43.2 & -57:44:17 & 0.0039 & 6 August 2018   & 1.94 &	6.48 & -0.273   &  0.188   &  Sy	   \\
1251-26  & ESO 507-G025           &12:51:31.8 & -26:27:07 & 0.0108 & 11 July 2018    & 1.89 &	5.11 & 0.103    &  0.045   &  LIER	   \\
1301-35  & ESO 381-G047           &13:01:05.4 & -35:37:00 & 0.0161 & 10 July 2018    & 1.91 &	5.42 & 0.042    &  0.131   &  LIER	   \\
1321-27  & NGC 5101               &13:21:46.2 & -27:25:50 & 0.0062 & 9 July 2018     & 1.96 &	6.94 & 0.361    &  0.505   &  LIER	   \\
1342-48b & NGC 5266               &13:43:02.1 & -48:10:10 & 0.0100 & 9 July 2018     & 1.98 &	7.27 & 0.212    &  0.275   &  LIER	   \\
1459-16  & NGC 5796               &14:59:24.1 & -16:37:26 & 0.0095 & 21 July 2017    & 1.96 &	6.82 & 0.380    &  0.0551  &  LINER	   \\
1503-13  & MCG-02-38-030          &15:03:00.2 & -13:16:58 & 0.0084 & 7 August 2018   & 1.82 &	3.71 & 0.177    &  0.572   &  LIER	   \\
1724-59  & ESO 138-G 024          &17:24:06.5 & -59:22:56 & 0.0096 & 9 August 2018   & 1.26 &	0.37 & -0.545   &  -0.396  &  SF	   \\
1758-53  & ESO 182-G001           &17:58:42.7 & -53:47:59 & 0.0119 & 8 August 2018   & 1.87 &	4.79 & 0.128    &  0.378   &  LIER	   \\
1914-62  & IC 4831                &19:14:43.8 & -62:16:21 & 0.0145 & 6 August 2018   & 1.82 &	3.84 & 0.112    &  0.533   &  LIER	   \\
1932-55  & NGC 6799               &19:32:16.5 & -55:54:29 & 0.0112 & 7 August  2018  & 1.83 &   3.98 & -0.056   &  0.231   &  LIER	   \\
1945-54  & IC 4889                &19:45:15.1 & -54:20:39 & 0.0086 & 20 July 2017    & 1.87 &	4.74 & 0.175    &  0.364   &  LIER	   \\
2013-37  & ESO 399-G025           &20:13:27.7 & -37:11:20 & 0.0085 & 12 July 2018    & 1.95 &	6.65 & -0.512   &  0.486   &  Sy	   \\
2015-21  & ESO 596-G012           &20:15:43.5 & -21:30:59 & 0.0196 & 11 July 2018    & 1.65 &	2.22 & -0.129   &  0.260   &  LIER	   \\
2018-16  & IC 1313                &20:18:43.6 & -16:56:45 & 0.0110 & 5 August 2018   & 1.89 &	5.18 & 0.066    &  0.440   &  LIER	   \\
2118-63  & IC 5096                &21:18:21.5 & -63:45:38 & 0.0105 & 6 August 2018   & 1.92 &	5.68 & 0.009    &  0.276   &  LIER	   \\
2153-37  & 2MASXJ21525329-3739110 &21:52:53.3 & -37:39:11 & 0.0149 & 25 August2018   & 1.80 &	3.38 & -0.187   &  0.817   &  Sy	   \\
2201-31  & ESO 466-G036           &22:01:20.4 & -31:31:47 & 0.0079 & 8 August 2018   & 1.35 &	0.54 & -0.920   &  0.413   &  SF	   \\
2224-03  & MCG-01-57-004          &22:23:39.1 & -03:25:54 & 0.0094 & 9 August 2018   & 1.63 &	2.05 & -0.62    &  -0.210  &  SF	   \\
2241-44  & IC 5240                &22:41:52.4 & -44:46:02 & 0.0059 & 9 July 2018     & 1.71 &	2.73 & -0.011   &  0.158   &  LIER	   \\
2243-64  & IC 5244                &22:44:13.7 & -64:02:36 & 0.0116 & 8 August 2018   & 1.95 &	6.65 & 0.0178   &  0.362   &  LIER	   \\
2334-04  & IC 5334                &23:34:36.4 & -04:32:03 & 0.0074 & 8 August 2018   & 1.82 &	3.68 & -0.167   &  0.319   &  Sy	   \\
  \hline&   
 \end{tabular}
  \begin{tablenotes}
      \small
      \item $^a$ Calculated using the narrow definition of \citet{balog99}.
      \item $^b$ Based on \citet{bruzual03} models using \citet{chabrier03} IMF and assuming Solar metallicity.
      \item $^c$ Emission line ratios measured from the integrated spectra composed of spaxels within a 1.5$\mathrm{^{\prime\prime}}$ radius from the centre of the galaxy. We refer to [O {\sc iii}]$\mathrm{\lambda~5007}$~\AA\ and [N {\sc ii}]$\mathrm{\lambda~6583}$~\AA\ as [O {\sc iii}] and [N {\sc ii}], respectively.
      \item $^d$ AGN classification for the sample following the definition of \citet{kauff03} and \citet{schawinski07}. Sy = Seyfert, LINER = low-ionisation nuclear emission-line region, LIER = low-ionisation emission-line region, SF = Star forming or H {\sc ii} regions. 
    \end{tablenotes}
  \end{threeparttable}
\end{table}
\end{landscape}

\subsection{Spectral Fitting}\label{sec:result}
To search for signs of recent SF and quantify the nuclear ionisation parameters of the galaxies in the sample, we extract emission line fluxes from the WiFes data cubes by following the procedure outlined below:
\begin{enumerate}
 \item The data is Voronoi binned using the procedure of \citet{cappellari03} to achieve adequate signal-to-noise (S/N) per bin based on the continuum S/N from 6000 to 6200 \AA\ range. The variance data cube given by the PyWiFeS pipeline was used to determine the noise. We choose a S/N target of 15.
 \item The stellar kinematics are then determined for each Voronoi bin spectra using the penalised pixel fitting \citep[\textsc{ppxf},][]{cappelari17, cappellari04} and the Indo-U.S. Library of Coude Feed Stellar Spectra templates \citep{valdes04}. Gas emission lines are also fitted simultaneously using \textsc{ppxf}. The [O {\sc iii}] and  [N {\sc ii}] doublets are fixed at theoretical flux ratio $\mathrm{\sim 3}$ \citep{cappelari17}.
 \item We measure nuclear spectra in a $3^{\prime\prime}$ diameter aperture, which at the mean redshift of the sample ($z=0.01$) corresponds to 0.6~kpc. Spaxels within a $1.5^{\prime\prime}$ radius from the centre of the galaxy are velocity shifted to the central spaxel using the determined stellar kinematics of the relevant Voronoi bin, and then stacked. The stacked spectrum is then fitted with \textsc{ppxf}. We refer to this integrated spectrum as the nuclear spectrum of a galaxy.
\end{enumerate}

We also measure the 4000 \AA\ break (D4000) of the nuclear spectrum of each galaxy by adopting the narrow definition of \citet{balog99}. We estimate stellar population ages by comparing the measured strength of D4000 with \citet{bruzual03} single stellar population models of Padova 1994 evolutionary tracks with \citet{chabrier03} IMF and assuming Solar metallicity. We note that the D4000 is not an accurate stellar age indicator due to its degeneracy with metallicity and the unknown star formation history, but it is sufficient for this work to separate galaxies with recent star-formation, whose light will be dominated by a young stellar population from relatively passive galaxies whose light will be dominated by old stars.

\section{Spectroscopic Results}\label{sec:results}

\begin{figure}
\includegraphics[width=\linewidth]{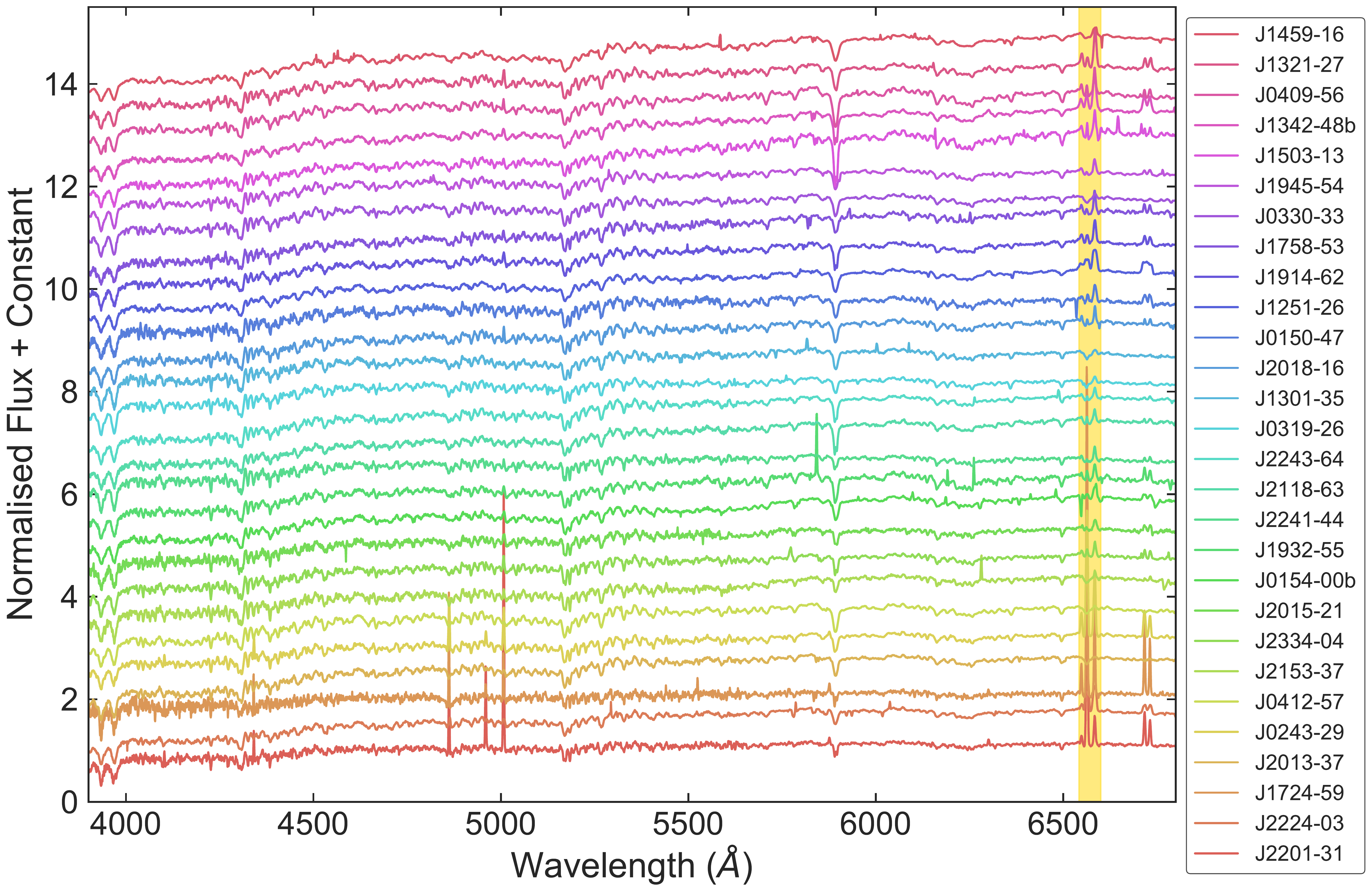}
\caption{The nuclear spectra shifted to the rest frame of the spectroscopic sample. The galaxies are ordered from bottom to top by increasing [N {\sc ii}]/H$\mathrm{\alpha}$ emission ratio. Balmer absorption lines and D4000 break are observed in 25 out of 28 of the galaxies, leading us to classify these spectra as being from ageing stellar populations. Most galaxies in the sample also exhibit a strong [N {\sc ii}] emission characteristic of Seyfert and LINER galaxies. }
\label{fig:spectra}
\end{figure}

In Figure \ref{fig:spectra}, we show the nuclear spectrum for the inner 1.5$\mathrm{^{\prime\prime}}$ of each galaxy with the H$\mathrm{\alpha}$ and [N {\sc ii}] wavelength range highlighted in yellow. Three out of 28 galaxies show evidence of central star formation, including H$\mathrm{\beta}$, [O {\sc ii}] and H$\mathrm{\alpha}$ emission lines and a relatively flat continuum with a weak 4000~\AA\ break of $\sim$ 1.4, which corresponds to an age of 1 Gyr. However, the vast majority of the galaxies (25/28) have a strong 4000~\AA\ break consistent with the centres of these galaxies having an old stellar population. We estimate the mean D4000 of these 25 galaxies to be 1.87 and the stellar population age to be $\sim$ 5 Gyr. While most of the galaxies have old stellar populations with strong 4000~\AA\ breaks, all the galaxies have significant nebular line emission. These 25 galaxies have [N {\sc ii}] to $\mathrm{H\alpha}$ flux ratios greater than one, which is consistent with emission from Seyferts (Sy), low-ionisation nuclear emission-line region (LINERs), or low-ionisation emission-line region (LIERs) galaxies. 

\begin{figure*}
\includegraphics[width=\linewidth]{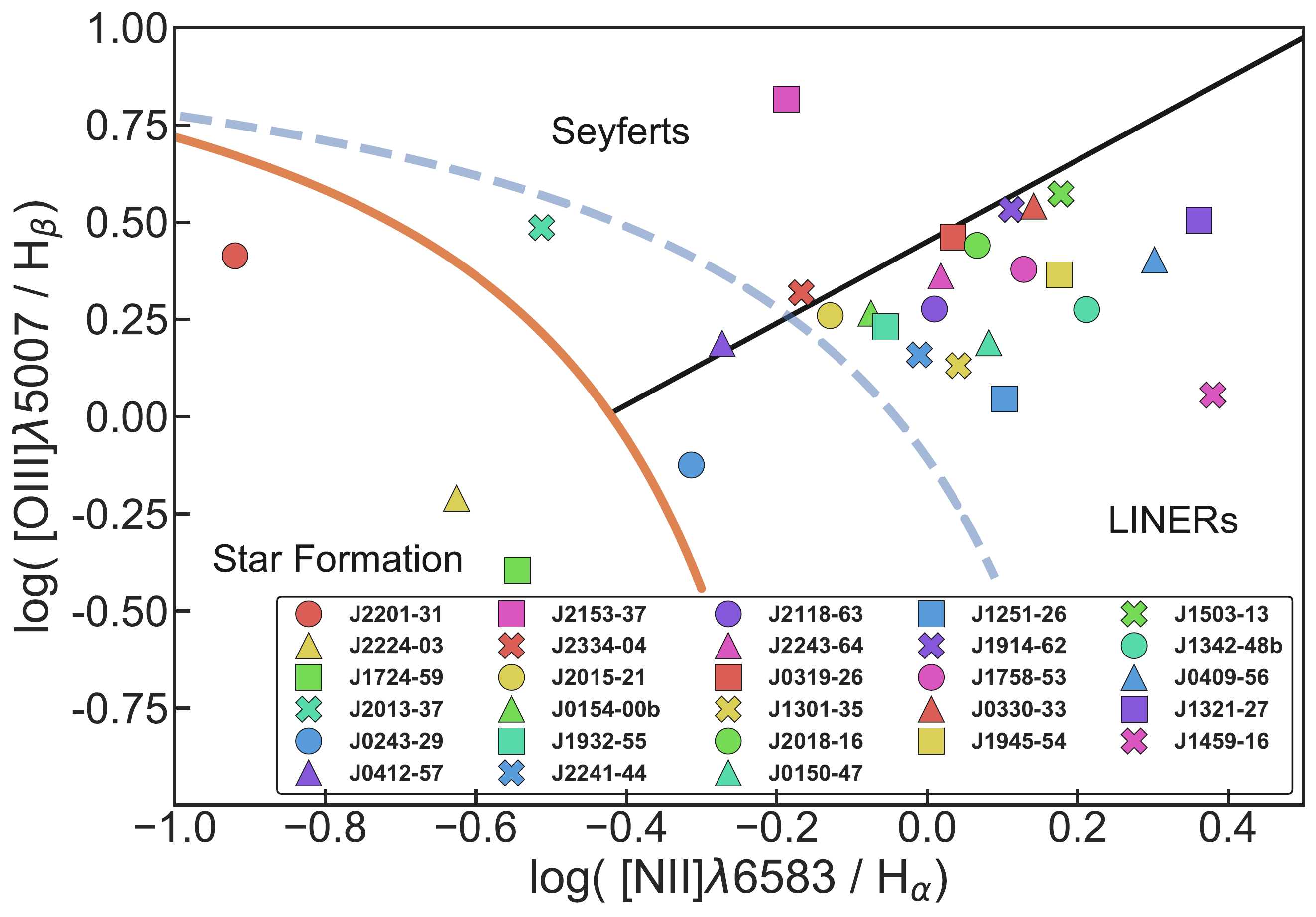}
\caption{BPT diagnostic diagram of the nuclear spectra (using a 3$\mathrm{^{\prime\prime}}$ diameter) for the spectroscopic sample. Each galaxy is represented by a different symbol to clearly display the classification of individual galaxies. The \citet{kauff03} and \citet{kewley01} classification lines, which are used to separate emission from H{\sc ii} regions from AGN-like emission, are in orange and blue respectively and the \citet{schawinski07} division line between Seyfert-like and LINER-like emission is in black. Though 3 galaxies exhibit star-forming like emission, the majority of the galaxies exhibit AGN-like emission in the central regions. Out of 25 galaxies with Seyfert-LINER like emission, 4 galaxies show Seyfert-like emission and 21 galaxies show LINER-like emission.}
\label{fig:BPT}
\end{figure*}

We classify the spectroscopic sample based on the Baldwin, Phillips and Terlevich diagram \citep[BPT; ][]{baldw81} using the nuclear spectra shown in Figure \ref{fig:BPT}. In this work, we adopt the demarcation line of \citet{kauff03} and \citet{schawinski07} to separate star-forming galaxies, Seyferts and LINERs/LIERs. We also distinguish between galaxies with nuclear low-ionisation emission and extended low-ionisation emission as LINERs and LIERs, respectively. Of the 28 galaxies observed, 3 galaxies are classified as star-forming, 4 are classified as Seyferts and 21 are classified as either LINERs or LIERs. We distinguish between LINERs and LIERs by considering the spaxels outside the nuclear region as further discussed in Section \ref{sec:lier}. Table \ref{tab:obs} lists the emission line ratios and AGN classification based on the central spectra. 

In Figure \ref{fig:6df} and in Appendix \ref{sec:appa}, we show examples of archival nuclear spectra of galaxies in the parent sample to compare with the WiFeS spectroscopic sample. We compiled 52 nuclear spectra from the 6dF Galaxy Survey \citep[6dFGS;][]{jones04, jones09}. Of the 52 galaxies, $\sim$ 35 (67\%) show evidence of  [N {\sc ii}] emission stronger than H$\mathrm{\alpha}$, indicative of LINERs or Seyferts. The total percentage of LINERs and Seyferts estimated from the archival 6dFGS spectra (67\%) may be an underestimate due to low signal-to-noise, so the high percentage of LINERs/LIERS in the WiFeS spectroscopic sample (75\%) may better reflect the true percentage in the parent sample. We do not provide a more quantitative analysis of the 6dFGS spectra because measurements of the emission lines are hindered by poor signal-to-noise and flux calibration.

\begin{figure}
\includegraphics[width=\linewidth]{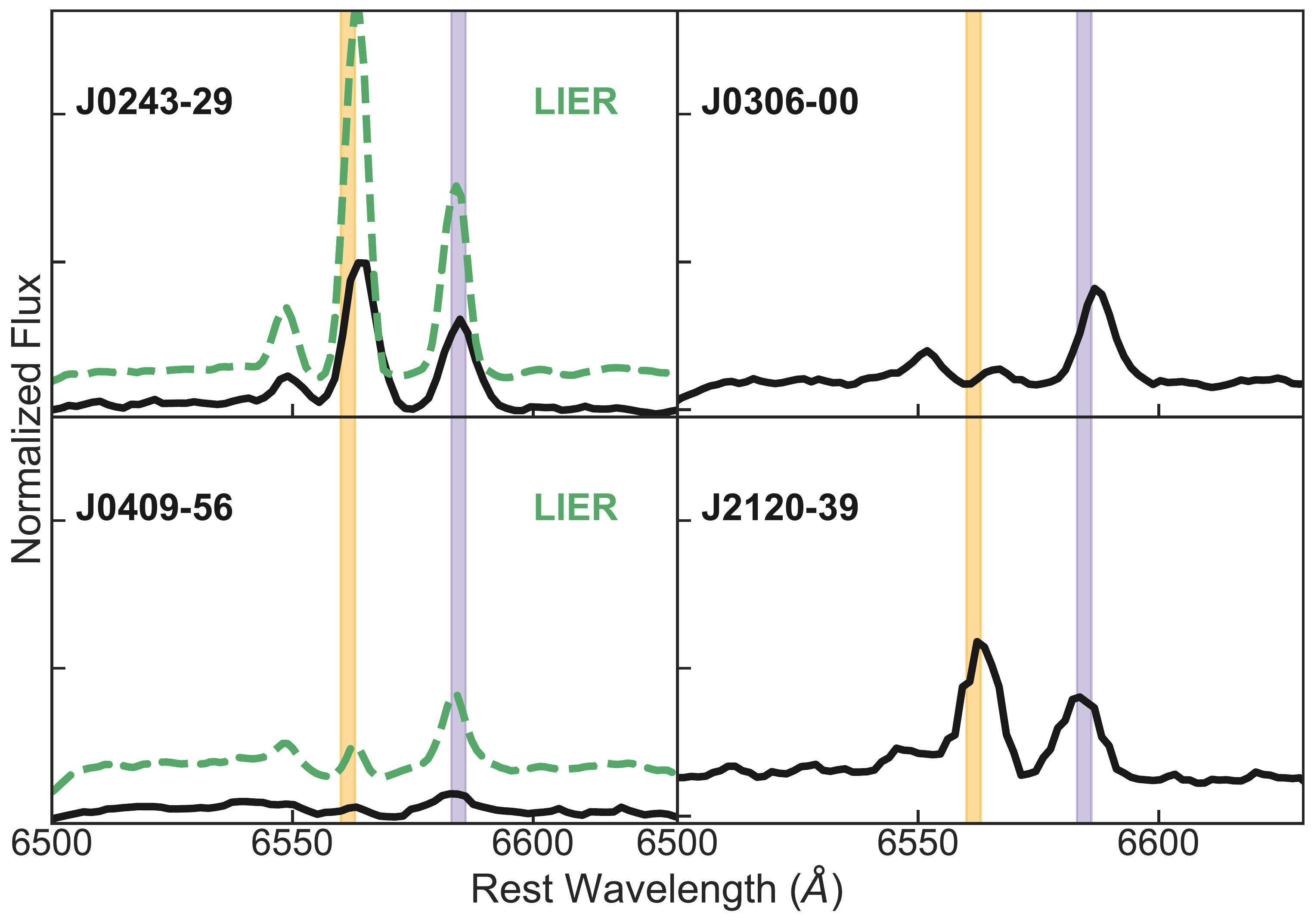}
\caption{Examples of archival 6dFGS spectra (black-solid line) shifted to rest frame of the nuclear region of the galaxies in the sample. The H$\mathrm{\alpha}$ and [N {\sc ii}] emission lines are highlighted in yellow and purple, respectively and the WiFeS nuclear spectum is shown in green when available. Out of the 52 galaxies in the sample that have archival 6dFGS nuclear spectra, $\sim$35 galaxies show [N {\sc ii}] emission that is comparable to or stronger than H$\mathrm{\alpha}$ emission, which is indicative of LINERs or AGNs rather than star formation.}
\label{fig:6df}
\end{figure}

\section{LINERs or LIERs?}\label{sec:lier}
To evaluate the extended emission of the galaxies and distinguish between LINERs and LIERs, we construct a spatially-resolved BPT diagram for each of the galaxies in the spectroscopic sample, which are shown in Figure \ref{fig:BPTresolved} and in Appendix \ref{sec:appb}. The line ratios are calculated for each Voronoi bin in the data-cube with a signal-to-noise (S/N) $>$ 3 in the relevant emission lines. Each Voronoi bin is coloured by the distance between the galaxy centre and the Voronoi bin centre. Bins within the central $1.5^{\prime\prime}$ of the galaxy are circled in black. We find that most of the emission line ratios of the outer Voronoi bins of the star-forming galaxies are classified as star forming. On the other hand, the excitation structure of Seyfert galaxies (i.e. galaxies where the central spectra are classified as Seyfert) appears to be more complex as the outer bins are either becoming more star-forming or having LINER-like emission.

To distinguish between LINERs and LIERs, we define emission from the nuclear region to be within a 3$\mathrm{^{\prime\prime}}$\footnote{At a the sample's mean redshift of 0.01, 3$\mathrm{^{\prime\prime}}$ corresponds to 0.6~kpc.} diameter aperture and extended emission to be outside that aperture. A galaxy is labelled as a LIER if at least 1 Voronoi bin located outside the nuclear region is classified as LINER-like emission. Galaxies with insufficient S/N outside the nuclear region or with no Voronoi bins outside the nuclear region exhibiting LINER-like emission are labelled as LINERs. Out of 21 galaxies that are LINERs or LIERs, 20 are classified as LIERs and 1 is classified as a LINER. Our principal results do not strongly depend on the threshold used for extended emission. If we use a 8$\mathrm{^{\prime\prime}}$ diameter to define the nuclear region then 18 galaxies are classified as LIERs and 3 galaxies are classified as LINERs. 

\begin{figure*}
\includegraphics[width=\linewidth]{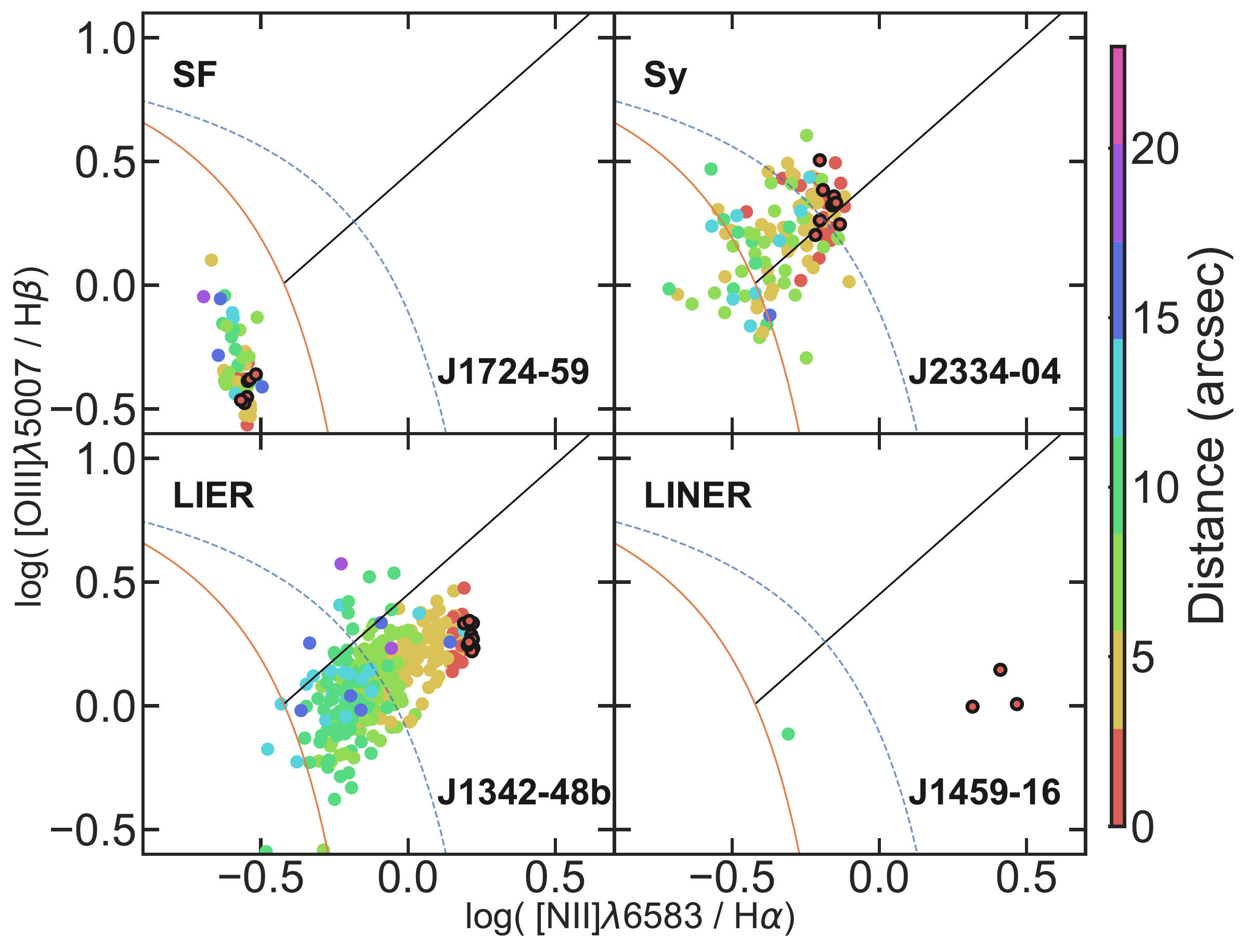}
\caption{Example spatially-resolved BPT emission line ratio diagrams of the different spectroscopic types for the spectroscopic sample. Similar to Figure \ref{fig:BPT}, classification lines are included on each subplot. The colour corresponds to distance between the centre of the Voronoi bin to the galaxy centre. Bins within the central 3$\mathrm{^{\prime\prime}}$ diameter of the galaxy are circled in black. Only bins with S/N $>$ 3 in all the relevant emission lines are shown. BPT diagrams for the remaining galaxies are given in Appendix \ref{sec:appb}. Out of the 21 galaxies to show central LINER emission in Figure \ref{fig:BPT}, 20 galaxies show emission that extends beyond $3^{\prime\prime}$, so we classify these galaxies as LIERs.}
\label{fig:BPTresolved}
\end{figure*}

\section{Discussion}\label{sec:dis}

Our initial motivation was to study the properties of galaxies that are in transition between the red sequence and blue cloud by studying a sample of \HI\ galaxies with little star formation and to identify possible quenching mechanisms that are at play. Using the \HI-WISE sample, we identified 91 \HI\ galaxies that show little to no star formation (sSFR $\mathrm{<~10^{-10.4} yr^{-1}}$) and obtained IFU follow-up observations of 28 \HI\ galaxies using the WiFeS IFS. Insights into how these galaxies might have been quenched is not immediately apparent, so we are saving such an analysis for a future work. However, the large fraction of LIERs in the spectroscopic sample is a standout result in itself. In this section, we will discuss the potential source of this LIER emission and why there is an unusually large fraction of LIERs in the spectroscopic sample.

The extended LIER emission that we observe in the spectroscopic sample is at odds with them being powered by low luminosity AGNs \citep[i.e.,][]{heckman80b, ho97, kauff03, kewley06}. For decades, low-luminosity AGNs were the preferred explanation for LINER emission, but recent studies have found many ``LINERs'' that are inconsistent with the AGN-ionisation hypothesis. Spatially resolved IFU observations, similar to our work, revealed that LINER-like emission is extended on kpc scales and that the radial emission-line surface brightness profiles of LINERs are shallower than the predicted r$^{-1}$ radial decrement for AGN photoionization \citep{sarzi10,yan12,singh13}. Therefore, it has been argued that the power source for extended LIER emission is not due to an AGN, but a source that is distributed throughout the galaxy.

Several authors have proposed that the source of LIER emission is from photoionisation by evolved stars, including post-asymptotic giant branch (pAGB) stars\footnote{pAGB stars includes all stages following the asymptotic giant branch including the hydrogen burning, white dwarf cooling and intermediate phases \citep{stanghellini00}.} \citep{binette94, stasinska08, cid10, cid11, yan12, belfiore16}. pAGB stars are hot ($\sim$ 30,000 K) enough to ionise the surrounding cool gas to produce LIER emission ratios and stellar population models have demonstrated that these stars become the main source of ionising photons once star formation has ceased and the young stars are dead \citep{binette94,sarzi10,johansson14,johansson16}. \citet{belfiore16} estimates the time from which pAGB stars become the dominant contributors to the ionising photon flux to be about $\sim$ 2 Gyrs after the last star forming event. Given that we have estimated an average age of stellar population for the LIERs to be $\sim$ 5 Gyrs, we conclude that pAGB stars to be the major contributor to the ionising photon flux in the galaxies and therefore the source of the LIER emission that we are observing.

Shocks caused by various astrophysical phenomena, such as galactic winds and mergers, may also explain LIERs \citep{monreal06, monreal10, farage10, rich10, rich11, rich15}. Gas kinematics may provide evidence for shock ionisation, and we will study the stellar and gas kinematics of the galaxies in a future paper. However, we note that deep optical imaging does not show compelling evidence for galaxy mergers. Thus, while shocks may explain some LIERs, for now our preferred explanation is pAGB stars.

Compared to previous studies, the high percentage of LIER galaxies in the spectroscopic sample is remarkable. At $z \sim 0$, for a stellar mass range between $10^{9}$ to $10^{12}$ \Msolar, $\sim$20\% of all galaxies are LINERs or LIERs, $\sim53\%$ are star-forming galaxies and the remainder include Seyferts, mergers and unclassified galaxies such as passive galaxies \citep{belfiore16}. For a stellar mass range between 10$^{10}$ to 10 $^{11.5}$ \Msolar, LINERs/LIERs make up $\sim$75\% of the spectroscopic sample. The percentage of LINERs/LIERs compared to the total galaxy population is found to increase with stellar mass, but only from $\sim 20\%$ to $\sim 35 \%$ for a $10^{10}$ to $10 ^{10.5}$ \Msolar\ stellar mass bin to a 10$^{11}$ to 10 $^{11.5}$ \Msolar\ stellar mass bin \citep{belfiore16}. This implies that the \HI\ content plus the low star formation rate of the sample is a common trait of LINERs/LIERs.

In the context of our sample selection and properties, the high percentage of LIERs in the spectroscopic sample may not be as surprising as it seems. In order to detect extended-LIER emission from a galaxy, the host galaxy must be old enough to have pAGB stars as the dominant ionising source and cold gas for the ultraviolet photons to ionise. LIER emission is not detected in star-forming galaxies as young stars are the dominant source of ionising photons \citep{binette94}. Meanwhile, not all old passive galaxies show LIER emission because they lack the cold gas required to produce the LIER signature spectrum \citep{belfiore17}. Though we do not have spatial information about \HI\ gas for the sample, we know that the galaxies have a \HI\ gas reservoir and presumably there is some amount of gas present in the central regions in order to produce the observed LIER emission. Follow-up interferometric \HI\ observations are needed to confirm the location of \HI\ gas relative to the observed LIER emission. Our broad picture, where LIERs result from evolved stellar populations and presence of cold gas, is broadly similar to that of \citet{belfiore17}, although they did not directly measure \HI\ gas content.

Previous studies have detected \HI\ gas in LINERs \citep{ haan08, gereb13, gereb15}.  \citet{haan08} found the mean \HI\ mass of seven LINERs to be $\mathrm{\sim10^{10}}$ \Msolar, which is comparable to the \HI\ masses of the most massive LIERs in the spectroscopic sample. \citet{gereb15} used stacking to detect \HI\ in $\mathrm{0 < z < 0.12}$ Seyferts and LINERs, and found a dependence on NUV-r colour, with only green valley objects containing detectable \HI\ gas while red objects did not. As these studies selected their AGNs and LINERs using nuclear spectra emission line ratios, it is not clear if they were selecting galaxies with extended LIER emission. However, these studies do indicate that \HI\ gas is common in green valley galaxies that have nebular emission lines and older stellar populations.

The characteristics of the LIERs in the spectroscopic sample are consistent with the inside-out quenching scenario \citep{munoz07,tacchella15,lin17,belfiore17}. The WiFeS spatially resolved data demonstrates that the centre of these galaxies are dominated by an older stellar population out to a radius of $\mathrm{\sim19^{\prime\prime}}$, or 3.8 kpc, implying that star formation has stopped, while deep optical imaging of 10 LIERs (i.e., HIPASSJ0243-29, HIPASSJ0319-26, HIPASSJ2241-44) show evidence of recent star formation on average 10 kpc (and sometimes up to 30 kpc) from the centre of the galaxies, consistent with the inside-out quenching scenario. 

\begin{figure}
\includegraphics[width=\linewidth]{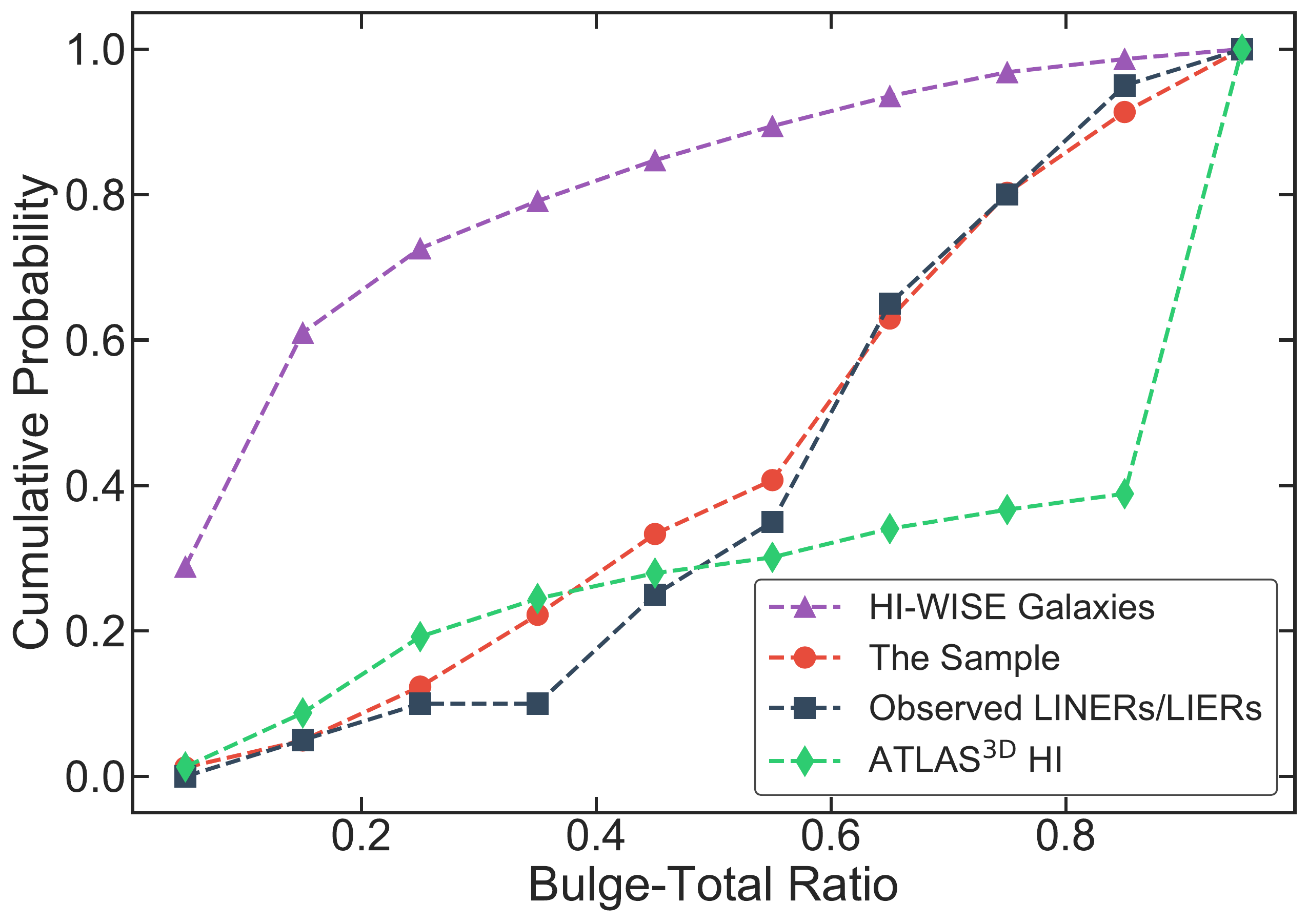}
\caption{The cumulative probability of bulge-to-total ratios of the sample and of the observed LINER/LIERs compared to the \HI-WISE sample \citep{parkash18} and $\mathrm{ATLAS^{3D}}$ \HI\ survey \citep{serra12, krajnovic13}. The bulge and disk light fractions are estimated using the axi-symmetric radial distribution of the W1 (3.4$\mathrm{\mu m}$) emission constructed for each galaxy, in which is fit and integrated a double Sersic profile consisting of the inner bulge and the extended disk components \citep[][2018, in preparation]{jarrett13}. We perform KS tests on B/T distributions for the \HI-WISE sample, the parent sample of 91 galaxies and the observed LIERs/LINERs. We find the parent sample of 91 galaxies and the LIERs/LINERs have comparable B/T distributions, but have significantly higher B/T ratios than the overall \HI-WISE galaxies. The B/T distribution for LIERs/LINERs is consistent with green valley galaxies, as they fall between the B/T distributions of star forming galaxies (i.e. \HI-WISE galaxies) and the bulge dominated passive galaxies (such as $\mathrm{ATLAS^{3D}}$). }
\label{fig:BT}
\end{figure}

Figure \ref{fig:BT} illustrates that LINERs/LIERs have a B/T distribution consistent with green valley galaxies that may be undergoing inside-out quenching \citep{schiminovich07, lin17}, as they fall between  the B/T distributions of star forming galaxies (i.e. \HI-WISE galaxies) and the bulge dominate passive galaxies (such as $\mathrm{ATLAS^{3D}}$). We preform a Kolmogorov-Smirnov test (KS test) on the B/T distributions for the \HI-WISE sample and the parent sample of 91 galaxies and find that we may reject the null hypothesis that the \HI-WISE sample and parent sample are drawn from the same distribution (KS statistic $=~0.62$). We find a similar result while performing the the KS test on the B/T distributions for the observed LIERs/LINERs sample and \HI-WISE sample (KS statistic $=~0.69$), emphasising that morphology of this sample differs from the star-forming galaxies of the \HI-WISE sample. That being said, the B/T ratio of the sample is not as large as the $\mathrm{ATLAS^{3D}}$ ETGs. One possible physical cause to inside-out quenching is AGN feedback \citep[e.g.,][]{nandra07,wake12,bluck14}, where an AGN heats or ejects cold gas, and hence terminates star formation in the bulge. We cannot rule out that the LINERs/LIERs host weak AGNs as we do not achieve the 100-200 pc resolution required to identify low luminosity AGNs \citep[e.g.,][]{Ho08, belfiore16} as we are observing in 1-2$\mathrm{^{\prime\prime}}$ seeing. Follow-up higher spatial resolution IFU observations are needed to confirm that AGN feedback is the cause of inside-out quenching.

\section{Summary and Conclusions}\label{sec:conclusion}
We have identified a sample of 91 local \HI\ galaxies ($\mathrm{10^{8.7}}$~\Msolar $<$ \Mh $<$ $\mathrm{10^{10.8}}$~\Msolar) with little to no star formation (sSFR $\mathrm{<~10^{-10.4}~yr^{-1}}$) and characterised their properties with IFU spectroscopy. While we have selected against star formation, all of the galaxies contain \HI\ and deep imaging reveals that most (and perhaps all) of the galaxies in the sample have low levels of star formation, often at relatively large radii ($\mathrm{\sim1^{\prime}}$, or $\sim$10 kpc). IFU observations of 28 galaxies reveals that majority of the galaxies (20 of 28 observed) are extended low ionisation emission region galaxies; LIERs. D4000 values imply that the LIERs have an average stellar population age of $\sim$ 5 Gyrs. The extended nebular line emission and dominant old stellar population are consistent with  pAGBs being responsible for the observed LIER emission. The high fraction of LIERs in the spectroscopic sample implies that presence the \HI\ gas combined with little to no star formation could be a precondition for LIER emission. 

\section*{Acknowledgements}
We would like to thank K. Lutz and R. D\v{z}ud\v{z}ar for the useful and insightful discussions and G. Hallenbeck for helpful comments that improved the paper. 

This publication makes use of data products from the Wide- field Infrared Survey Explorer, which is a joint project of the University of California, Los Angeles, and the Jet Propulsion Laboratory/California Institute of Technology, funded by the National Aeronautics and Space Administration. Funding for the SDSS III has been provided by the Alfred P. Sloan Foundation, the U.S. Department of Energy Office of Science, and the Participating Institutions. The observational data presented in this paper were collected as part of the CGS (http://cgs.obs.carnegiescience.edu), using facilities at Las Campanas Observatory, Carnegie Institution for Science. The data were reduced independently from those presented in \citet{ho11}. Based on observations made with the NASA Galaxy Evolution Explorer. GALEX is operated for NASA by the California Institute of Technology under NASA contract NAS5-98034.Based on photographic data obtained using The UK Schmidt Telescope. The UK Schmidt Telescope was operated by the Royal Observatory Edinburgh, with funding from the UK Science and Engineering Research Council, until 1988 June, and thereafter by the Anglo-Australian Observatory. Original plate material is copyright (c) of the Royal Observatory Edinburgh and the Anglo-Australian Observatory. The plates were processed into the present compressed digital form with their permission. The Digitized Sky Survey was produced at the Space Telescope Science Institute under US Government grant NAG W-2166.

This research has made use of the NASA/IPAC Extragalactic Database (NED) which is operated by the Jet Propulsion Laboratory, California Institute of Technology, under contract with the National Aeronautics and Space Administration and the HyperLeda database (http://leda.univ-lyon1.fr). This research made use of Astropy, a community-developed core Python package for Astronomy (Astropy Collaboration, 2018). 

The Pan-STARRS1 Surveys (PS1) and the PS1 public science archive have been made possible through contributions by the Institute for Astronomy, the University of Hawaii, the Pan-STARRS Project Office, the Max-Planck Society and its participating institutes, the Max Planck Institute for Astronomy, Heidelberg and the Max Planck Institute for Extraterrestrial Physics, Garching, The Johns Hopkins University, Durham University, the University of Edinburgh, the Queen's University Belfast, the Harvard-Smithsonian Center for Astrophysics, the Las Cumbres Observatory Global Telescope Network Incorporated, the National Central University of Taiwan, the Space Telescope Science Institute, the National Aeronautics and Space Administration under Grant No. NNX08AR22G issued through the Planetary Science Division of the NASA Science Mission Directorate, the National Science Foundation Grant No. AST-1238877, the University of Maryland, Eotvos Lorand University (ELTE), the Los Alamos National Laboratory, and the Gordon and Betty Moore Foundation.

This project used public archival data from the Dark Energy Survey (DES). Funding for the DES Projects has been provided by the U.S. Department of Energy, the U.S. National Science Foundation, the Ministry of Science and Education of Spain, the Science and Technology Facilities Council of the United Kingdom, the Higher Education Funding Council for England, the National Center for Supercomputing Applications at the University of Illinois at Urbana-Champaign, the Kavli Institute of Cosmological Physics at the University of Chicago, the Center for Cosmology and Astro-Particle Physics at the Ohio State University, the Mitchell Institute for Fundamental Physics and Astronomy at Texas A\&M University, Financiadora de Estudos e Projetos, Funda{\c c}{\~a}o Carlos Chagas Filho de Amparo {\`a} Pesquisa do Estado do Rio de Janeiro, Conselho Nacional de Desenvolvimento Cient{\'i}fico e Tecnol{\'o}gico and the Minist{\'e}rio da Ci{\^e}ncia, Tecnologia e Inova{\c c}{\~a}o, the Deutsche Forschungsgemeinschaft, and the Collaborating Institutions in the Dark Energy Survey. The Collaborating Institutions are Argonne National Laboratory, the University of California at Santa Cruz, the University of Cambridge, Centro de Investigaciones Energ{\'e}ticas, Medioambientales y Tecnol{\'o}gicas-Madrid, the University of Chicago, University College London, the DES-Brazil Consortium, the University of Edinburgh, the Eidgen{\"o}ssische Technische Hochschule (ETH) Z{\"u}rich,  Fermi National Accelerator Laboratory, the University of Illinois at Urbana-Champaign, the Institut de Ci{\`e}ncies de l'Espai (IEEC/CSIC), the Institut de F{\'i}sica d'Altes Energies, Lawrence Berkeley National Laboratory, the Ludwig-Maximilians Universit{\"a}t M{\"u}nchen and the associated Excellence Cluster Universe, the University of Michigan, the National Optical Astronomy Observatory, the University of Nottingham, The Ohio State University, the OzDES Membership Consortium, the University of Pennsylvania, the University of Portsmouth, SLAC National Accelerator Laboratory, Stanford University, the University of Sussex, and Texas A\&M University. Based in part on observations at Cerro Tololo Inter-American Observatory, National Optical Astronomy Observatory, which is operated by the Association of Universities for Research in Astronomy (AURA) under a cooperative agreement with the National Science Foundation.


\bibliographystyle{mnras}
\bibliography{Thesis.bib} 

\appendix

\section{Sample}\label{sec:sample_app}
Table \ref{table:full_sample} list the properties of the \HI\ galaxies in the sample as reported in \citet{parkash18}. Luminosity distance (D$_L$) were compiled from (in order of preference) Cosmicflows-3 \citep{tully16} and HICAT redshift. \HI\ mass were estimated using the published integrated 21 cm flux ($F_{\textup{H{\textsc i}}}$) from HICAT \citep{meyer04}:
\begin{eqnarray}
\textup{M}_{\textup{H{\textsc i}}} [\textup{M}_\odot] = \frac{2.356\times10^5}{1+z} \times  D_L^2\times \textup{F}_{\textup{H{\textsc i}}}[\textup{Jy km s}^{-1}].
\end{eqnarray}
Stellar mass were estimated following the GAMA-derived stellar mass-to-light ($M_\ast/L_{W1_\odot}$) ratio relation of \citet{cluver14}:
\begin{eqnarray} \label{eq:GAMA}
\log_{10}M_\ast/ L_{\mathrm{W1}_{\mathrm{Sun}}}= -1.96(\mathrm{W1} - \mathrm{W2}) - 0.03,
\end{eqnarray}
which depends on the ``in-band" luminosity relative to the Sun,
\begin{eqnarray}
L_{\mathrm{W1}_{\mathrm{Sun}}} = 10^{-0.4(M - M_{\mathrm{Sun}}) },
\end{eqnarray}
where $M$ is the absolute W1 magnitude and $M_{\mathrm{Sun}}$ $=$ 3.24 \citep{jarrett13}. SFR were derived from the estimated W3 emission flux from the interstellar medium (ISM), W3PAH \citep{cluver17}. In \citet{parkash18}, we estimate Balmer-decrement-corrected H$\alpha$ ($L_{H_{\alpha},\mathrm{Corr}}$) from the W3PAH flux using the prescription from \citep{brown17} and derives SFRs scaling the \citet{kennicutt98} calibration to a \citet{chabrier03} IMF:
\begin{eqnarray}
 \log{L_{W3_{\mathrm{PAH}}}}\; [\textup{erg s}^{-1}] = (40.79 \pm 0.06) + (1.27 \pm 0.04) \times \\ (\log{L_{H_{\alpha},\mathrm{Corr}}} \;[\textup{erg s}^{-1}] -40),
\end{eqnarray}
and
\begin{eqnarray}
SFR\; [M_\odot\: \textup{yr}^{-1}]=( 4.6\times10^{-42} ) \times L_{H_{\alpha},\mathrm{Corr}}\;[\textup{erg s}^{-1}] .
\end{eqnarray}
As a consequence of using the W3PAH flux to estimate SFR, 16 of the 91 \HI\ galaxies in the sample do not have a SFR measurement since the measured W3 flux of these galaxies is mostly dominated by evolved stellar population.

T-types reported on Table \ref{table:full_sample} are extracted from \citet{bonne15} catalogue. Visual notes on the galaxy features or orientation based on optical CGS, DES or Pan-STARRS imaging are also listed such as bars (B), rings (R), star formation in the outer regions of the galaxy (SF), dust lanes (DL), or if the galaxy was edge-on (EO), disturbed or irregular (D), or a merging system (M).
\onecolumn
\begin{longtable}{cccccccccc}
\caption{The sample}\label{table:full_sample}\\
HIPASS ID & RA & DEC & D$_L^a$ & z & T-type & Notes$^b$ & log \Mh & log \Ms & log SFR \\ 
& (J2000) & (J2000) & (Mpc) & &  & & (${\rm M_\odot}$) & (${\rm M_\odot}$) &(${\rm M_\odot}$ yr$^{-1}$)  \\
\hline
\endfirsthead
\multicolumn{10}{c}%
{\tablename\ \thetable\ -- \textit{Continued from previous page}} \\
\hline
HIPASS ID & RA & DEC & D$_L$ & z & T-type & Notes & \Mh & \Ms & SFR \\ 
\hline
\endhead
\hline \multicolumn{10}{r}{\textit{Continued on next page}} \\
\endfoot
\hline \\
\multicolumn{10}{l}{$^a$ Luminosity Distance. }\\
\multicolumn{10}{l}{$^b$ Visual notes based on CGS, DES, Pan-STARRS, DSS and/or GALEX images (if available). B: bar, R: ring,  }\\
\multicolumn{10}{l}{SF: evidence of star formation in the outer region of the galaxy, EO: edge-on, DL: dust lane, D: disturbed or irregular, }\\
\multicolumn{10}{l}{M: merged or merging, P: passive, ETG: early-type galaxy. }\\
\endlastfoot
0010-59 & 00:10:41 & -59:41:41 & 100.89 & 0.02308 & 4.0 & B,R,SF  & 10.00 & 10.51 & -.25 \\
0058-15 & 00:58:20 & -15:23:31 & 71.59 & 0.01646 & 2.0 & B,R,SF   & 10.35 & 10.62 & -.18 \\
0150-47 & 01:50:29 & -47:09:57 & 87.79 & 0.02013 & 2.0 & R,SF     & 10.07 & 10.73 & -.92 \\
0154-00b & 01:54:41 & -00:08:36 & 79.65 & 0.01829 & -5.0 & R,SF   & 9.73 & 10.78 &    \\
0227-01 & 02:27:37 & -01:09:21 & 30.05 & 0.00696 & 0.0 & B,R      & 8.71 & 11.45 & -.65 \\
0238-01 & 02:38:12 & -01:19:08 & 34.52 & 0.00799 & 1.0 & B,R,SF   & 9.47 & 10.61 & -.47 \\
0243-29 & 02:43:44 & -29:00:11 & 26.06 & 0.00604 & 0.0 & B,R,SF   & 9.57 & 10.69 & -.54 \\
0244-09 & 02:44:41 & -09:07:35 & 100.44 & 0.02298 & -5.0 & P,ETG  & 9.86 & 10.35 & -.67   \\
0248-31 & 02:48:41 & -31:32:10 & 89.52 & 0.02052 & 3.0 & EO,DL    & 10.24 & 10.83 & -.03 \\
0306-00 & 03:06:52 & -00:47:40 & 30.74 & 0.00712 & 0.0 & B,R,SF   & 9.44 & 10.55 & -.99 \\
0315-03 & 03:15:52 & -03:37:48 & 56.56 & 0.01304 & 5.0 & D        & 9.86 & 10.59 & -.45 \\
0317-41 & 03:17:19 & -41:06:29 & 10.54 & 0.00245 & 0.0 & B,R,SF   & 9.36 & 11.12 & -.41 \\
0319-26 & 03:19:51 & -26:03:38 & 22.55 & 0.00523 & 0.0 & B.R,SF   & 9.25 & 10.79 & -.47 \\
0330-33 & 03:31:08 & -33:37:42 & 20.43 & 0.00474 & 1.0 & B,R,SF,DL& 9.11 & 10.93 & -.16 \\
0335-24 & 03:35:01 & -24:55:59 & 31.48 & 0.00729 & 1.0 & B,SF     & 10.21 & 11.14 & -.19 \\
0339-22 & 03:39:11 & -22:23:20 & 25.28 & 0.00586 & -4.0 & B,SF   & 8.69 & 10.15 &   \\
0342-06 & 03:42:10 & -06:45:55 & 70.71 & 0.01626 & 0.0 & B,R      & 10.12 & 10.46 & -.10 \\
0352-66 & 03:53:34 & -66:01:05 & 77.67 & 0.01784 & -5.0 & P,ETG   & 9.90 & 11.47 &   \\
0354-36a & 03:54:28 & -35:58:02 & 21.47 & 0.00498 & -4.0 & R,P,ETG& 9.02 & 10.57 &   \\
0409-56 & 04:09:52 & -56:07:06 & 22.38 & 0.00519 & -2.0 & B,R     & 9.93 & 10.89 & -.98 \\
0412-57 & 04:12:43 & -57:44:17 & 20.04 & 0.00465 & -2.0 & B,R,SF  & 8.93 & 11.05 & -.51 \\
0423-09 & 04:24:02 & -09:23:41 & 141.68 & 0.03219 & 5.0 & B,SF    & 10.18 & 11.05 &   \\
0436-03 & 04:36:37 & -03:11:20 & 71.1 & 0.01635 & 1.0 & R,SF      & 10.47 & 11.24 & .08   \\
0439-07 & 04:39:31 & -07:05:51 & 68.24 & 0.0157 &  & B,R          & 10.237& 10.80 & -1.65 \\
0441-01a & 04:41:37 & -01:48:32 & 47.64 & 0.011 & -2.0 & P,ETG    & 9.54 & 10.97 & -.21 \\
0442-20 & 04:42:15 & -20:26:05 & 22.63 & 0.00525 & 3.0 & B,R,SF   & 9.24 & 10.35 & -.41 \\
0448-57 & 04:48:57 & -57:39:33 & 100.93 & 0.02309 & 1.0 & B,R,SF  & 10.26 & 10.59 & -.07 \\
0514-61 & 05:14:36 & -61:28:54 & 65.56 & 0.01509 & 0.0 & SF       & 10.26 & 10.95 & -.45 \\
0531-30 & 05:31:28 & -30:36:08 & 135.51 & 0.03082 &  & B,R,SF     & 10.11 & 10.99 & -.52 \\
0550-06 & 05:50:24 & -06:36:59 & 94.89 & 0.02173 & 1.0 & EO,DL    & 10.47 & 10.84 & -.03 \\
0621-27b & 06:21:40 & -27:14:02 & 27.27 & 0.00632 & 0.0 & B,R,SF  & 9.73 & 11.28 & .12   \\
0622-13 & 06:22:01 & -13:53:27 & 46.2 & 0.01067 & -2.0 & P,ETG    & 9.55 & 10.02 & -1.11 \\
0626-63 & 06:26:12 & -63:45:16 & 81.16 & 0.01863 & -2.0 &         & 10.02 & 10.83 & -.22 \\
0646-26a & 06:46:22 & -26:06:31 & 35.91 & 0.00831 & 0.0 & M       & 9.82 & 10.63 & -.49 \\
0755-52 & 07:55:52 & -52:18:24 & 17.83 & 0.00414 & -2.0 & B       & 9.12 & 10.76 &   \\
0911-14 & 09:11:28 & -14:49:00 & 33.91 & 0.00785 & 0.0 & B,R,SF   & 9.59 & 10.95 & -.18 \\
0913-63 & 09:13:32 & -63:37:35 & 23.45 & 0.00544 & 2.0 & SF       & 9.55 & 10.55 & -.06 \\
0930-14 & 09:30:53 & -14:44:09 & 37.13 & 0.00859 & -2.0 & R       & 9.59 & 10.78 & -1.32 \\
0955-13 & 09:55:01 & -13:05:09 & 146.6 & 0.03328 &  & EO          & 9.87 & 10.94 &   \\
1001-34 & 10:01:30 & -34:06:45 & 49.4 & 0.0114 & 0.0 &            & 9.70 & 10.92 & -.23 \\
1041-28 & 10:41:42 & -28:46:47 & 74.13 & 0.01704 & 5.0 & SF       & 10.43 & 10.76 & -.31 \\
1041-46 & 10:41:26 & -46:17:14 & 107.16 & 0.02449 & 0.0 & EO      & 9.95 & 10.85 &   \\
1051-17 & 10:51:37 & -17:07:29 & 88.16 & 0.02021 & 2.0 & B,SF     & 10.46 & 10.19 & -.19 \\
1127-28 & 11:27:05 & -28:58:49 & 29.05 & 0.00673 & 0.0 & R,SF     & 9.27 & 10.56 & -.72 \\
1159-28 & 11:59:17 & -28:54:18 & 33.96 & 0.00786 & -2.0 & R       & 9.65 & 10.15 &   \\
1215-35 & 12:15:34 & -35:37:47 & 36.83 & 0.00852 & 3.0 & SF,DL    & 8.88 & 11.04 & -.02 \\
1234-40 & 12:34:43 & -40:18:02 & 48.9 & 0.01129 & 0.0 & B,R       & 9.55 & 10.78 & -.35 \\
1243-41 & 12:43:31 & -41:21:43 & 29.79 & 0.0069 & -2.0 & P,ETG    & 9.08 & 10.71 & -.35 \\
1245-06 & 12:45:41 & -06:04:09 & 26.14 & 0.00606 & 9.0 & SB       & 9.42 & 10.04 & -.83 \\
1251-26 & 12:51:32 & -26:27:07 & 40.35 & 0.00933 & -4.0 & DL,SF   & 10.40 & 11.21 & -.20 \\
1253-08 & 12:53:36 & -08:38:19 & 55.51 & 0.0128 & -5.0 & B,SF     & 9.88 & 10.91 & -1.13 \\
1256-46 & 12:56:41 & -46:55:34 & 76.56 & 0.01759 & -2.0 &         & 9.72 & 10.90 & -.25 \\
1301-35 & 13:01:05 & -35:37:00 & 74.1 & 0.01703 & -2.0 & R        & 9.99 & 10.93 & -.75 \\
1304-30 & 13:04:17 & -30:31:34 & 48.51 & 0.0112 & -5.0 & P,ETG    & 9.46 & 11.73 &   \\
1304-41 & 13:04:05 & -41:24:41 & 41.27 & 0.00954 & 4.0 & B,R,SF   & 10.18 & 11.18 & -.02 \\
1308-23 & 13:08:04 & -23:47:49 & 47.99 & 0.01108 & -4.0 & P,ETG   & 9.96 & 10.93 & -1.50 \\
1309-37 & 13:09:22 & -37:10:27 & 56.04 & 0.01292 & -5.0 & P,ETG   & 9.81 & 10.44 & -2.39 \\
1309-51b & 13:09:35 & -51:58:07 & 59.85 & 0.01379 & -2.0 & P,ETG  & 10.04 & 11.39 & -.22 \\
1313-16 & 13:13:12 & -16:07:50 & 44.15 & 0.0102 & -2.0 & P,ETG    & 9.57 & 10.11 &   \\
1320-21 & 13:20:17 & -21:49:38 & 19.13 & 0.00444 & -2.0 & EO,DL   & 9.98 & 10.99 & -.57 \\ 
1321-27 & 13:21:46 & -27:25:50 & 17.36 & 0.00403 & 0.0 & B,R,SF   & 9.50 & 10.88 & -.19 \\
1324-37 & 13:24:46 & -37:40:56 & 24.97 & 0.00579 & 1.0 & R,SF     & 9.19 & 10.66 & -.36 \\
1342-48b & 13:43:02 & -48:10:10 & 25.93 & 0.00601 & -4.0 & DL,ETG & 9.72 & 11.06 & -.18 \\
1416-75 & 14:16:59 & -75:38:50 & 39.18 & 0.00906 & 0.0 & P        & 9.67 & 10.52 & -.42 \\
1459-16 & 14:59:24 & -16:37:26 & 44.36 & 0.01025 & -5.0 & P,ETG   & 9.30 & 11.32 & -.28 \\
1503-13 & 15:03:00 & -13:16:57 & 32.96 & 0.00763 & 1.0 & EO       & 9.35 & 10.31 & -.63 \\
1515-36 & 15:15:56 & -36:45:13 & 65.29 & 0.01503 & -2.0 & ETG,P   & 10.04 & 10.80 &   \\
1617-34 & 16:17:40 & -34:21:57 & 30.05 & 0.00696 & 1.0 &          & 9.15 & 10.62 & -.43 \\
1723-80b & 17:26:04 & -80:11:42 & 72.69 & 0.01671 & -2.0 &        & 9.99 & 10.99 & -.53 \\
1724-59 & 17:24:07 & -59:22:56 & 41.88 & 0.00968 & 5.0 & EO       & 9.65 & 10.10 & -.66 \\
1758-53 & 17:58:43 & -53:47:59 & 51.35 & 0.01185 & -2.0 &         & 9.29 & 10.89 & -.30 \\
1823-61 & 18:22:17 & -61:52:13 & 61.65 & 0.0142 & 4.0 & D         & 9.85 & 10.07 &   \\
1914-62 & 19:14:44 & -62:16:21 & 62.26 & 0.01434 & 2.0 & DL,SF    & 10.42 & 11.26 & .26   \\
1932-55 & 19:32:17 & -55:54:28 & 52.01 & 0.012 & -5.0 & P, ETG    & 9.91 & 10.97 & -.35 \\
1933-58 & 19:33:19 & -58:06:51 & 61.13 & 0.01408 & 2.0 & DL,EO    & 10.16 & 11.44 & .11   \\
1945-54 & 19:45:15 & -54:20:39 & 28.83 & 0.00668 & -5.0 & ETG,P   & 9.10 & 10.96 & -.78 \\
2013-37 & 20:13:28 & -37:11:20 & 66.35 & 0.01527 & 1.0 & R        & 10.14 & 11.04 &   \\
2015-21 & 20:15:44 & -21:30:58 & 82.0 & 0.01882 & -2.0 & SF       & 10.36 & 10.76 & -.41 \\
2018-16 & 20:18:44 & -16:56:45 & 44.1 & 0.01019 & 1.0 & B,R,SF    & 10.08 & 10.74 & -.29 \\
2043-80 & 20:43:58 & -80:00:03 & 37.65 & 0.00871 & -2.0 &         & 9.40 & 11.16 & -.30 \\
2118-63 & 21:18:22 & -63:45:38 & 43.49 & 0.01005 & 4.0 & EO,DL    & 9.55 & 10.96 & .05   \\
2120-39 & 21:20:08 & -39:46:07 & 47.96 & 0.01107 & 1.0 & B,R      & 9.61 & 10.62 & -.26 \\
2122-36 & 20:13:28 & -37:11:20 & 66.35 & 0.01527 & 1.0 & D        & 10.19 & 11.04 &   \\
2123-69 & 21:23:29 & -69:41:05 & 154.82 & 0.0351 & 5.0 & B,R      & 10.80 & 10.90 &   \\
2144-75 & 21:44:16 & -75:06:41 & 33.48 & 0.00775 & 1.0 & B,R,SF   & 9.98 & 10.98 & -.03 \\
2153-37 & 21:52:53 & -37:39:11 & 60.99 & 0.01405 & -4.0 & P,ETG   & 9.79 & 10.19 & -.77 \\
2201-31 & 22:01:20 & -31:31:46 & 61.65 & 0.0142 & 1.0 & SF        & 9.93 & 10.33 & -.69 \\
2224-03 & 22:23:39 & -03:25:54 & 35.65 & 0.00825 & -2.0 & B,R     & 9.46 & 10.66 & -.57 \\
2241-44 & 22:41:52 & -44:46:02 & 34.04 & 0.00788 & 1.0 & B,R,SF   & 9.52 & 10.82 & -.15 \\
2243-64 & 22:44:14 & -64:02:36 & 48.82 & 0.01127 & 3.0 & EO,DL    & 9.88 & 10.94 & -.13 \\
2334-04 & 23:34:36 & -04:32:04 & 45.5 & 0.01051 & 3.0 & EO        & 9.98 & 10.58 & -.63 \\ \hline
\end{longtable}

\twocolumn
\section{Archival Spectrum}\label{sec:appa}
\begin{figure*}
\includegraphics[width=0.9\linewidth]{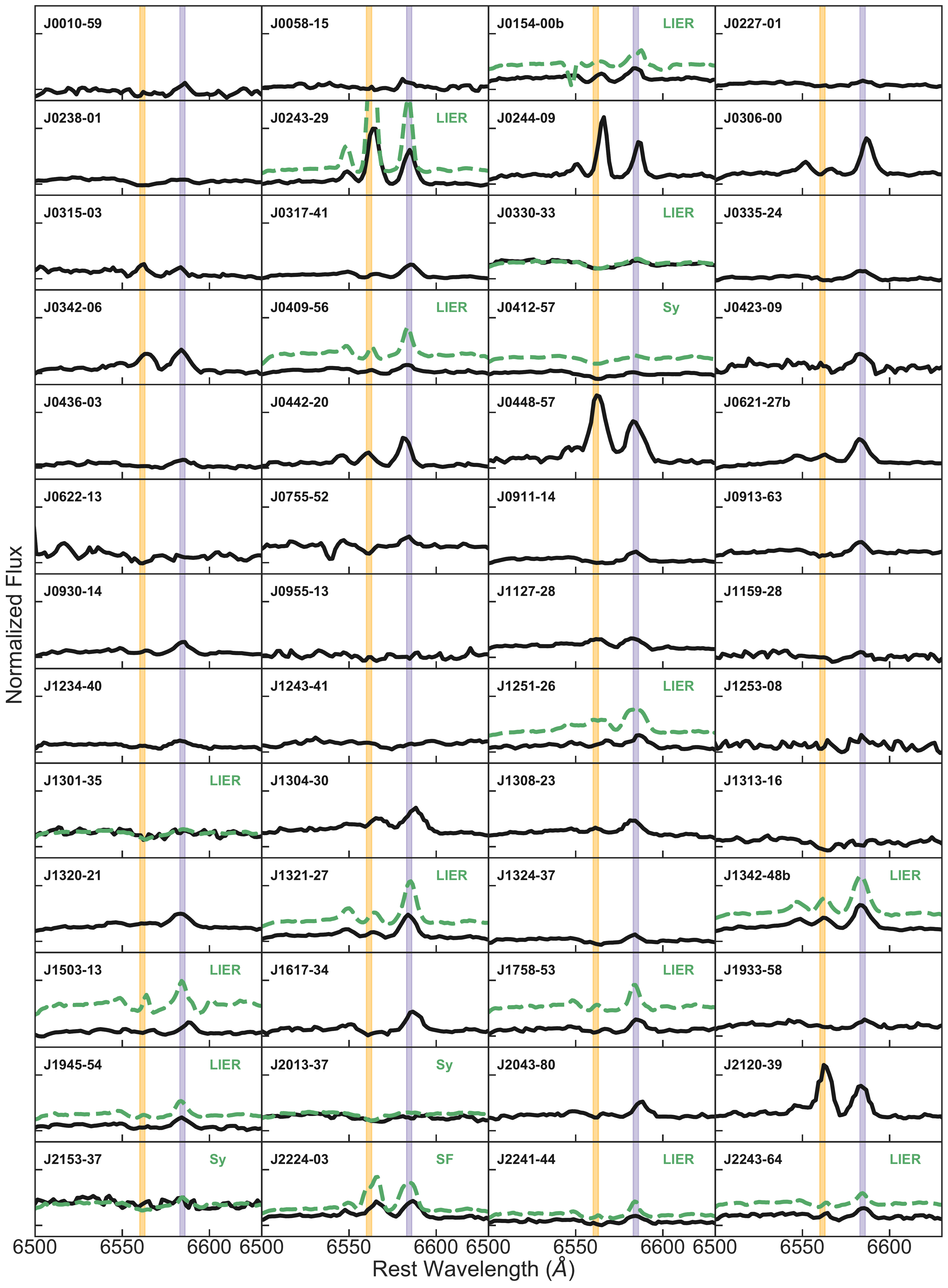}
\caption{6dFGS spectra (black-solid line) shifted to rest frame of the nuclear region of the galaxies in the sample. The H$\mathrm{\alpha}$ and [N {\sc ii}] emission lines are highlighted in yellow and purple, respectively and the WiFeS nuclear spectra is show in green if available.}
\label{fig:6df_long}
\end{figure*}

Fig. \ref{fig:6df_long} presents the archival nuclear spectrum for the galaxies in the parent sample. These spectrum are extracted from the 6dFGS. 

\section{BPT DIAGRAMS}\label{sec:appb}
\begin{figure*}
\includegraphics[width=\linewidth]{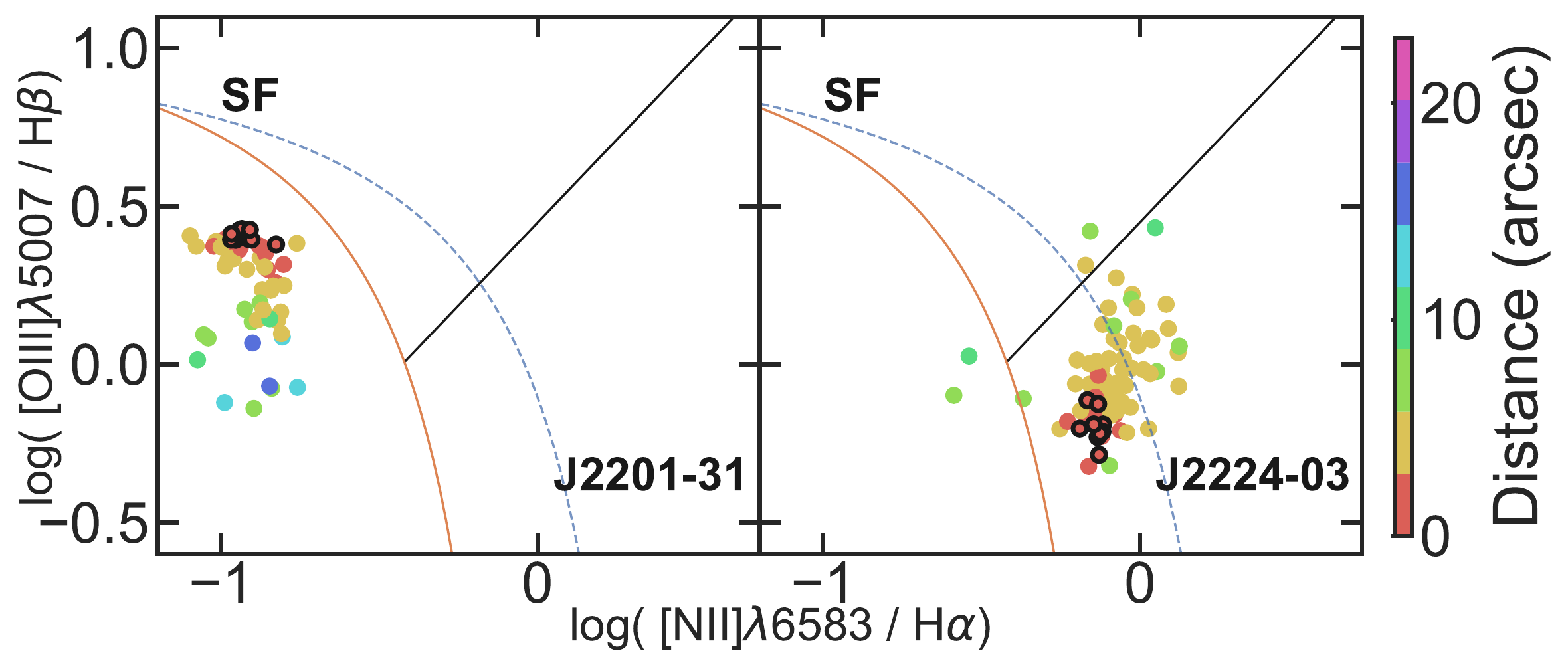}
\caption{Spaxel-by-spaxel BPT diagrams of the remaining spectroscopically classified star-forming galaxies in the sample. Classification lines are included on each subplot. The colour corresponds to distance between the centre of the Voronoi bin to the galaxy centre. Bins within the central 3$\mathrm{^{\prime\prime}}$ diameter of the galaxy are circled in black. Only bins with S/N $>$ 3 in all the relevant emission lines are shown. }
\label{fig:BPTresolved_sf}
\end{figure*}

\begin{figure*}
\includegraphics[width=\linewidth]{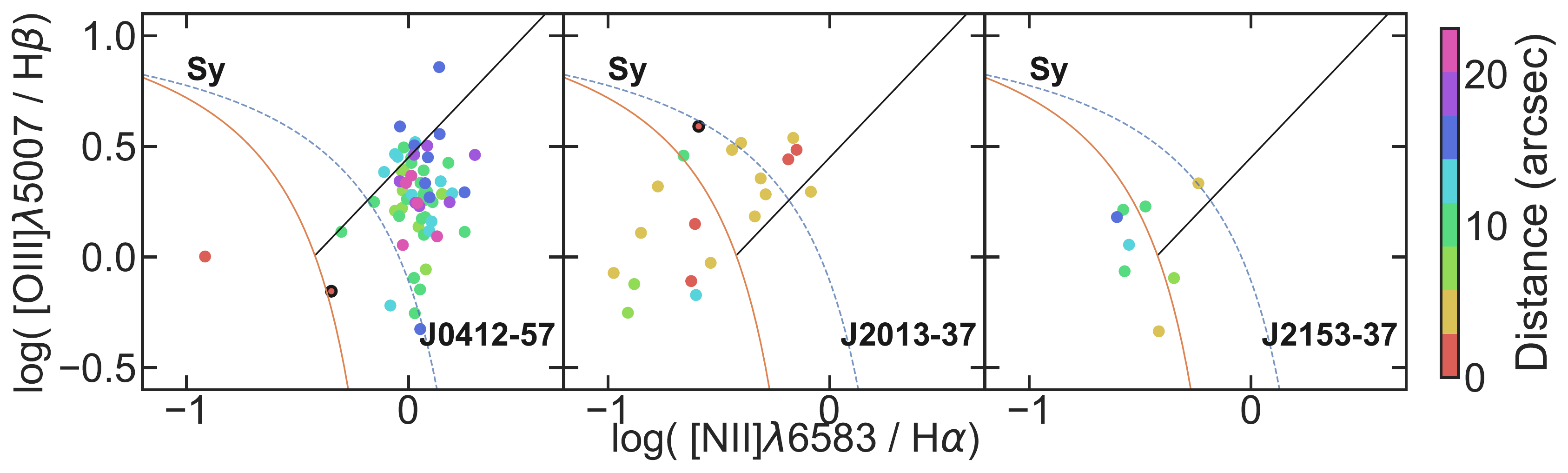}
\caption{Spaxel-by-spaxel BPT diagrams of the remaining spectroscopically classified Seyferts in the sample. Classification lines are included on each subplot. The colour corresponds to distance between the centre of the Voronoi bin to the galaxy centre. Bins within the central 3$\mathrm{^{\prime\prime}}$ diameter of the galaxy are circled in black. Only bins with S/N $>$ 3 in all the relevant emission lines are shown. }
\label{fig:BPTresolved_SY}
\end{figure*}

\begin{figure*}
\includegraphics[width=\linewidth]{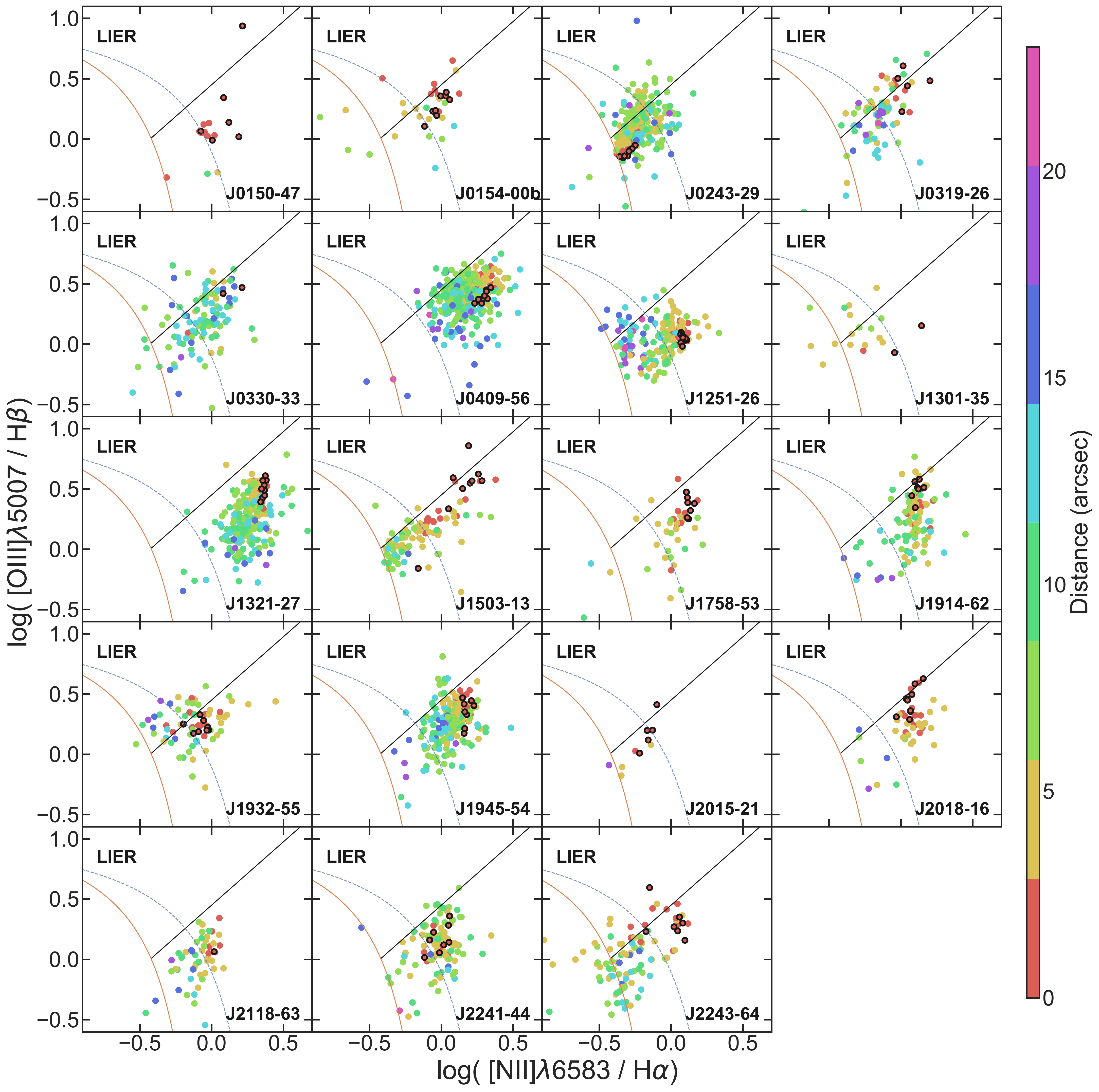}
\caption{Spaxel-by-spaxel BPT diagrams of the remaining spectroscopically classified LIERs in the sample. Classification lines are included on each subplot. The colour corresponds to distance between the centre of the Voronoi bin to the galaxy centre. Bins within the central 3$\mathrm{^{\prime\prime}}$ diameter of the galaxy are circled in black. Only bins with S/N $>$ 3 in all the relevant emission lines are shown. }
\label{fig:BPTresolved_lier}
\end{figure*}

Fig. \ref{fig:BPTresolved_sf}, \ref{fig:BPTresolved_SY}, and \ref{fig:BPTresolved_lier} presents the remaining spaxel-by-spaxel BPT diagrams for the spectroscopic sample. The galaxies are divided into spectroscopic types.

\bsp	
\label{lastpage}
\end{document}